\documentclass[fleqn,usenatbib]{mnras}

\usepackage{newtxtext,newtxmath}
\usepackage[T1]{fontenc}
\usepackage{ae,aecompl}
\usepackage{graphicx}	
\usepackage{amsmath}	
\usepackage{amssymb}

\pdfoutput=1

\title[Grain Alignment Efficiency in Dust Polarization]{Effects of Grain Alignment Efficiency on Synthetic Dust Polarization Observations of Molecular Clouds}

\author[P. K. King et al.]{
Patrick K. King,$^{1,2}$\thanks{E-mail: pkk4hu@virginia.edu}
Che-Yu Chen$^{1}$,
L. M. Fissel$^{3}$,
and Zhi-Yun Li$^{1}$
\\
$^{1}$Department of Astronomy, University of Virginia, Charlottesville, VA, 22904\\
$^{2}$Lawrence Livermore National Laboratory, Livermore, CA, 94550 \\
$^{3}$National Radio Astronomy Observatory, Charlottesville, VA, 22903
}

\date{Accepted XXX. Received YYY; in original form ZZZ}

\pubyear{2019}

\begin{document}
\label{firstpage}
\pagerange{\pageref{firstpage}--\pageref{lastpage}}
\maketitle

\begin{abstract}
It is well known that the polarized continuum emission from magnetically aligned dust grains is determined to a large extent by local magnetic field structure. However, the observed significant anticorrelation between polarization fraction and column density may be strongly affected, perhaps even dominated by variations in grain alignment efficiency with local conditions, in contrast to standard assumptions of a spatially homogeneous grain alignment efficiency. Here we introduce a generic way to incorporate heterogeneous grain alignment into synthetic polarization observations of molecular clouds, through a simple model where the grain alignment efficiency depends on the local gas density as a power-law. We justify the model using results derived from radiative torque alignment theory. The effects of power-law heterogeneous alignment models on synthetic observations of simulated molecular clouds are presented. We find that the polarization fraction-column density correlation can be brought into agreement with observationally determined values through heterogeneous alignment, though there remains degeneracy with the relative strength of cloud-scale magnetized turbulence and the mean magnetic field orientation relative to the observer. We also find that the dispersion in polarization angles-polarization fraction correlation remains robustly correlated despite the simultaneous changes to both observables in the presence of heterogeneous alignment. 
\end{abstract}


\section{Introduction} \label{section:intro}

Molecular Clouds (MCs) are the dense regions of the interstellar medium (ISM) where all known star formation occurs \citep{BT, M1}. Since the outcome of star formation depends on a balance between collapse under gas self-gravity and support provided by gas thermal/kinetic pressure and magnetic fields \citep{M1}, understanding magnetic field strength and organisation (collectively \textit{magnetic structure}) within MCs is one of the keys to understanding the star formation process. Despite its value, magnetic structure is difficult to ascertain, being relatively inaccessible observationally \citep{HGMZ}. Dust grains within the ISM are known to align with the local magnetic field \citep[and references therein]{ALV}, and polarized thermal emission offers a means to measure projected plane-of-sky magnetic structure. The advent of high-sensitivity submillimeter polarimetry that can map dust polarization over a large area \citep{PXIX, FBP} has led to an explosion of interest in leveraging these observational capabilities to ascertain details of the magnetic structure within MCs, especially as alternative methods, such as Zeeman measurements, are extremely challenging \citep{C1}. However, one cannot simply read off magnetic structure from these polarimetric observations, as the polarization state observed by the telescope is determined both by grain alignment/emission processes \citep{ALV} and the magnetic structure over the line-of-sight \citep{FP1}. This challenge has spurred the development of concurrent simulation and modelling efforts and data analysis techniques alongside these expanded telescopic capabilities in order to provide the necessary links between observed properties and the true, three-dimensional magnetic structure. 

The combination of these techniques could be brought under a unified framework termed \textit{observational magnetohydrodynamics}. The beginnings of this framework could be traced to the seminal work of \citet{CF1}, which provided a means of estimating the local plane-of-sky magnetic field strength from polarized starlight measurements (and later expanded to polarized dust emission measurements). Though the applicability of this technique is necessarily limited \citep{OSG1}, it serves as a conceptual template for newer techniques obtained through hard-won developments in MHD and star formation theory \citep{ES1, M1}, the development of precision far IR/submillimeter observational techniques for studying MCs \citep{HGMZ, H4, BT}, and the properties and magnetic alignment of dust grains in the interstellar medium \citep{ALV}. The maturation of high-sensitivity, high-angular resolution submillimeter polarimetry has resulted in a sea change in the sophistication of observational MHD, as detailed mapping leads in turn to high quality statistical characterization, using techniques such as the histogram of relative orientations (HRO) \citep{SOLER13,CKL,SOLER} and the projected Rayleigh statistic \citep{JOW}; observational population characteristics, and later, probability density function (PDF) estimation, and its relation to physical conditions \citep{FG1,PJP,PXIX,FBP,KFCL}; bispectrum techniques \citep{BISPECTRUM}; and the exploitation of the physics of polarized emission to constrain large-scale magnetic field properties \citep{CKLFM}, among others.

Within this landscape of observational MHD techniques, the statistical characterization of observable populations and joint correlations has the most direct overlap between synthetic and real observations. The BLASTPol observations revealed these polarization properties of the nearby MC Vela C to unprecedented statistical detail, and showed that Vela C displays significant anti-correlation between column density and polarization fraction \citep{FBP}, a property that had already been reported to be ubiquitous in MCs by other observations \citep{MWF, PFG, ALVES14, J15, PXIX, ALV}. Leveraging this new statistical power, \citet[hereafter Paper I]{KFCL} provided a comparison to the BLASTPol data with synthetic observations under the assumption of spatially uniform (or homogeneous) grain alignment to examine the role of magnetic structure on polarization observations. Though those simulated clouds cover a wide range of column densities and polarization levels, it was not possible to obtain column density-polarization fraction correlations consistent with these observations using a homogeneous alignment prescription, which suggests that this correlation cannot arise purely due to naturally-formed magnetized gas structure alone, and that grain alignment physics likely plays a role in producing these trends. 

Progress in grain alignment theory \citep{ALV} has led to the implementation of grain alignment models in full radiative transfer modelling of numerical simulations \citep{CL1, BCLK, SILCC}, but these calculations remain uncertain because of systematic uncertainties associated with the external radiation field conditions and grain population characteristics. We present a simple prescription for heterogeneous grain alignment in the form of a power-law in gas density, motivated by calculations made using results from the more exact work on radiative torque alignment physics in \citet{CL1} and \citet{BCLK}, while still remaining agnostic with respect to some details of grain alignment microphysics which are subject to systematic uncertainties. We then apply this model to an expanded set of colliding flow simulations to build on the results of Paper I. We use our synthetic observations to explore some of the basic effects of heterogeneous alignment on polarization observations. These synthetic observations are significantly affected by the orientation of the mean magnetic field and by the turbulence and magnetization conditions of the MC. This demonstrates that magnetic structure and grain alignment physics are both key factors controlling observational characteristics of MCs. Finally, though some ambiguity exists between the effects of heterogeneous alignment and magnetic structure, we show that it is possible to reproduce column density-polarization fraction correlations consistent with both the BLASTPol observations of Vela C, while simultaneously satisfying other observational constraints obtained through the calculation of the dispersion in polarization angles, $\mathcal{S}$. Additionally, we show that the filamentary structures seen in maps of $\mathcal{S}$ (e.g. \citet{PXX}) are uniquely affected by heterogeneous alignment models, and that the bulk of the $\mathcal{S}$ population is an important characterization of magnetic structure in MCs. 

Our paper is organized as follows. In Section \ref{section:methods} we introduce the colliding flow simulations used in our study (Section \ref{section:sims}) and the basic principles underlying our synthetic polarimetric observations and the framework for modifying the grain alignment prescription (Section \ref{section:hetalign}). In Section \ref{section:PLDM} we introduce our simple power-law heterogeneous alignment model and motivate it using some simple calculations based on existing radiative torque alignment modelling. We present the generic results of applying our heterogeneous alignment model in Section \ref{section:plresults}, focusing on the distributions of the polarization fraction and dispersion in polarization angles alone in Section \ref{section:pl1D}, the correlations with column density in Section \ref{section:plCD}, and the joint polarimetric correlation in Section \ref{section:plSp}. Finally we conclude in Section \ref{section:conc}. 

\section{Methods} \label{section:methods}

\subsection{Numerical Simulations} \label{section:sims}

\begin{figure}
\includegraphics[width=\columnwidth]{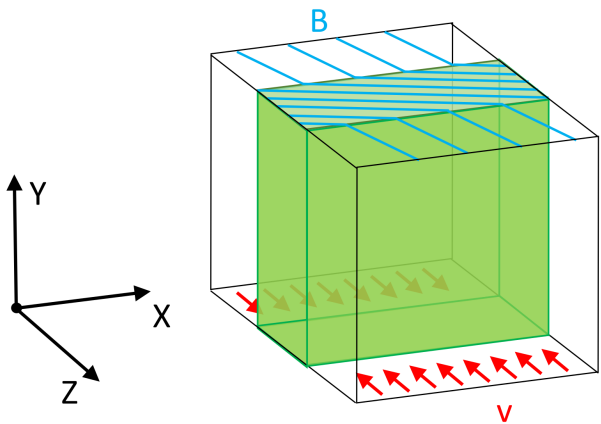}
\vspace{-2mm}
\caption{Geometry of the \textsc{Athena} simulations with notation conventions for the Cartesian coordinates annotated, reproduced from \citet{KFCL}. Converging flows produce a sheet-like post-shock region, in which the initially oblique magnetic field is compressed and bent.}
\label{fig:geometry}
\end{figure}

\begin{table*}
	\centering
	\caption{Some properties of the four \textsc{Athena} simulations, including: the initial number density, $n_0$; the simulation box side length, $L$; the pre-shock inflow Mach number $\mathcal{M}_{s,0}$; the amplitude of the turbulent velocity perturbation, $\sigma_{v,0}$; the initial magnetic field strength, $B_0$; the root-mean-square average post-shock velocity, $v_{rms,ps}$; the post-shock sonic and Alfv\'{e}n Mach numbers $\mathcal{M}_{s,ps}$ and $\mathcal{M}_{A,ps}$; the post shock plasma $\beta$; and the applied scaling transformation $\lambda_{scale}$, described in Paper I. The density and magnetic field strength values are scaled using the listed scaling transformation. Models B and D (annotated by an asterisk) are the same simulations as Models A and B, respectively, in Paper I.}
	\label{tab:simstats}
	\begin{tabular}{lcccccccccccc} %
		\hline
		 Simulation & $n_0$ & $L$ & $\mathcal{M}_{s,0}$ & $\sigma_{v,0}$ & $B_0$ & $\langle n_{ps}\rangle$ & $v_{rms,ps}$ & $\mathcal{M}_{s,ps}$ & $\mathcal{M}_{A,ps}$ & $\beta_{ps}$  & $\lambda_{scale}$ \\
		\hline
        Model A  & 50.0 cm$^{-3}$ & 20.0 pc & ~5.0 & 1.40 km/s & ~5.0 $\mu$G & 411 cm$^{-3}$ & 2.15 km/s & 10.75 & 1.39 & 0.03 & 1.0~~~~~ \\
		Model B* & 50.0 cm$^{-3}$ & 10.0 pc & 10.0 & 0.72 km/s & ~5.0 $\mu$G & 397 cm$^{-3}$ & 1.87 km/s & ~9.35 & 1.45 & 0.05 & 1.0~~~~~ \\
        Model C  & 50.0 cm$^{-3}$ & ~5.0 pc & 10.0 & 0.36 km/s & ~2.0 $\mu$G & 858 cm$^{-3}$ & 0.99 km/s & ~4.95 & 1.33 & 0.14 & 1.0~~~~~ \\
        Model D* & 50.0 cm$^{-3}$ & 4.47 pc & 10.0 & 0.14 km/s & 2.23 $\mu$G & 971 cm$^{-3}$ & 0.76 km/s & ~3.80 & 1.04 & 0.15 & 0.2236   \\
		\hline
	\end{tabular}
\end{table*}

In Paper I, two colliding flow simulations were chosen for study, which were similar to simulations discussed in \citet{CO1, CO2}. For illustration purposes, will use the same models used in that work, augmented by two new simulations of the same set-up, each with different initial conditions so as to achieve slightly different conditions in the post-shock region. This set of simulations has first appeared in \citet{CKLFM}, and the general characteristics of these simulations were discussed in more detail in Section 2.2 of Paper I and Section 4.1 of \citet{CKLFM}; however, we provide brief overview here. These are 3D, ideal MHD simulations with gravity, computed using the \textsc{Athena} code \citep{ATHENA}. An isothermal equation of state is adopted with a sound speed of 0.2 km/s. A supersonic converging flow is set up, and turbulence is seeded by perturbing the velocity field of the inflow with a Gaussian random distribution with a prescribed power spectrum $v_k^2 \propto k^{-4}$ \citep{GO1}. Initially, the magnetic field is constant and oriented 20$^{\circ}$ to the inflow direction, which results in a bent and compressed mean magnetic field (with turbulent perturbations) oriented mainly in the plane of the post-shock region. This set-up is used to simulate a cloud formation scenario, with the resulting post-shock region serving as the model MC. The resulting simulation geometry (neglecting turbulent perturbations) is shown in Figure \ref{fig:geometry}. For each simulation a single timestep is chosen for synthetic observations, corresponding to when the (unscaled) maximum density of the gas has reached $n_{max} = 10^7$ cm$^{-3}$. Maps of gas structures and polarization can be found in Figure 9 of \citet{CKLFM}.

Some of the initial and post-shock properties of the simulations are provided in Table \ref{tab:simstats}. As the post-shock region in each simulation is taken to be our modelled cloud, the post-shock average sonic and Alfv\'{e}n Mach numbers taken in tandem serve as a proxy for the relative importance of turbulent motions to the magnetic field. However, it must be noted that either is not a perfect proxy for relative magnetic disorder: the sonic Mach number does not contain information about the magnetic field, and even super-Alfv\'{e}nic motion oriented along the magnetic field lines will not change the magnetic field orientation. The models are arranged in descending order in post-shock Mach number/Alfv\'{e}n Mach number.\footnote{As subsequently noted, difficulties in defining the post-shock region can complicate the measures reported in Table \ref{tab:simstats}, which shows a slightly higher $\mathcal{M}_{A,ps}$ for Model B than Model A.} (Model A has a higher post-shock sonic Mach number than its corresponding inflow Mach number; the inflow Mach number neglects the turbulent velocity contributions and represents only the bulk velocity of the inflow, so this high post-shock sonic Mach number is sensible.) Progressing from Model D to Model A results in conditions which are more turbulent (relative to the sound speed) and with more magnetic field tangling and disorder. The post-shock Alfv\'{e}n Mach number is limited by the physics of oblique magnetized shocks \citep{CLKF}, and cannot be freely chosen; as a result, our simulations are limited to a trans-Alfv\'{e}nic regime. Models B and D correspond to Models A and B in Paper I, respectively; Model A in this work corresponds to the strongest turbulence, and Model C is intermediate between the models used in that work. A scaling transformation of the type described in Section 2.2.2 of Paper I is applied such that all have the same initial density $n_0$ to ensure that the density structures within the simulations are consistent.\footnote{This is a different scaling choice than chosen in Paper I, which was chosen to make the simulations consistent in length scale. The application of this scaling reduces the maximum density in Model D by a factor of 20, but given the extremely low frequency of such voxels, it will not significantly affect the bulk properties of the simulated MC.} There is some difficulty in unambiguously identifying the post-shock region in these simulations on density or velocity thresholding alone, as the turbulence present can drastically reduce the contrast between pre- and post-shock material, even in the relatively strong shocks in this study. As a result, the numerical value of the post-shock properties listed in Table \ref{tab:simstats} for Models B and D differ slightly from the numbers reported in Paper I due to the updated selection criteria for post-shock material. 

We select only lines-of-sight we are confident do not contain only pre-shock material, as described in Section 2.3 of Paper I. We choose to eliminate these lines-of-sight at the post-integration, pixel level as opposed to the pre-integration, voxel level for several reasons. Pragmatically, for all of our simulations except Model D, the turbulent perturbations are strong enough for the pre-shock and post-shock region to have significant overlap in their respective density populations. This can be seen below in Figure \ref{fig:densitypdfs}. As a result, there is no unambiguous density criterion that separates pre-shock and post-shock regions cleanly; for similar reasons (i.e. turbulence), this is not improved even when augmenting that condition with a velocity criterion to attempt to pick out the stagnated shock. This ambiguity is worsened if clear post-shock voxels are excluded according to this criterion, creating zero-emission holes in the region of interest. Allowing pre-shock voxels to contribute to emission also mimics real observational conditions, in which the ability of the observer to subtract diffuse emission is limited. By restricting our analysis at the pixel level we also follow observational practice by excluding low S/N regions from the observational sample. Nevertheless, our choice does introduce the vulnerability to contaminating the polarization signal from regions of substantially different magnetohydrodynamical conditions from the post-shock region, and if the contributions from high-density voxels are suppressed, this signal may be amplified. 

\subsection{Synthetic Polarimetric Observations} \label{section:hetalign}

The calculation of the synthetic Stokes parameters from simulation data is outlined in Paper I, which follows existing convention (e.g., \citealt{LD1,FP1,K1,PXX,CKL}). Under the assumption that polarized emission is in optically thin conditions, the gas is isothermal, and that the dust grains are in thermal equilibrium with the gas, the synthetic Stokes parameters in a Cartesian coordinate system are:

\begin{equation}
I ~ = \int n \left(1 - p_0\left(\frac{B_x^2 + B_y^2}{B^2} - \frac{2}{3}\right) \right) ds,
\label{I}
\end{equation}
\begin{equation}
Q = \int p_0~ n \left(\frac{B_y^2 - B_x^2}{B^2}\right) ds,
\label{Q}
\end{equation}
\begin{equation}
U = \int p_0~ n \left(\frac{2B_x B_y}{B^2}\right) ds.
\label{U}
\end{equation}

\noindent Here, $n$ is the gas number density, $\mathbf{B} = \{B_x,B_y,B_z\}$ is the local magnetic field, $p_0$ is the polarization efficiency, and $s$ is the distance along the line-of-sight. The validity of these approximations in computing the Stokes parameters was recently evaluated against polarized radiative transfer in \citet{SILCC}, and the authors demonstrated that it is highly accurate, with position angle deviations on the order of $\sim 5^{\circ}$. 

Adopting the assumption of homogeneous alignment, in which the polarization efficiency is the same everywhere, allows one to take $p_0$ out of the integrals and adopt a single uniform value, which in an observational context is treated as a property of the observed object in question. For this reason, adopting homogeneous alignment is tantamount to considering only the effects of magnetic structure in the synthetic observations. Using these forms, the polarization fraction is 

\begin{equation}
p = \frac{\sqrt{Q^2 + U^2}}{I}
\label{p}
\end{equation}

\noindent and the polarization angle is 

\begin{equation}
\chi = \frac{1}{2}\arctan(U,Q),
\label{ch}
\end{equation}

\noindent The dispersion in polarization angles \citep{H1,PXIX} may be computed at any position $\mathbf{x}$ and choice in lag distance $\delta$:

\begin{equation}
\mathcal{S}^2(\mathbf{x},\delta) = \frac{1}{N}\sum_{i=1}^N \Delta \chi^2(\mathbf{x},\mathbf{x}_i).
\label{S}
\end{equation}

\noindent where $\{\mathbf{x}_i\}$ is the set of all points located a distance $\delta$ from $\mathbf{x}$, and $\Delta \chi$ is the angular difference (consistent with $\chi$ being an orientation indistinguishable from a $\pi$ rotation) between $\chi$ at $\mathbf{x}$ and $\mathbf{x}_i$.

In Paper I, synthetic observations were presented at both the simulation pixel scale (corresponding to a resolution of the scaled box length divided by 512) and convolved with a Gaussian beam to explore the effects of beam convolution on the synthetic observations. The overall effect of beam convolution was found to primarily wash out extreme values in the data relative to the mean, i.e. extremely low or high values, for each observable. This was not found to alter the means of the observable distributions or substantially alter the principal components or correlation coefficients in the joint correlations. For this reason we will present the synthetic observations in this work only at the pixel scale of the simulations to show the full range of their distributions. 

Most of the statistical techniques used here are outlined in Section 2.4 of Paper I, including kernel density estimation (KDE) for estimating univariate and bivariate probability distributions; principal component analysis (PCA) for interpreting bivariate correlations; and geometric point statistics and correlation coefficients for characterization. The synthetic polarization observations were implemented using \textsc{python} code written using \textsc{numpy} \citep{NP}, \textsc{SciPy} \citep{SP}, \textsc{Numba} \citep{NUMBA}, and the \textsc{yt} package \citep{YT}. Our plots were produced using the \textsc{matplotlib} plotting library \citep{MPL}.

\section{Heterogeneous Grain Alignment} \label{section:PLDM}

\subsection{Power-Law Depolarization Models} \label{subsection:PLDM}

\begin{figure*}
\centering 
\includegraphics[width=\textwidth]{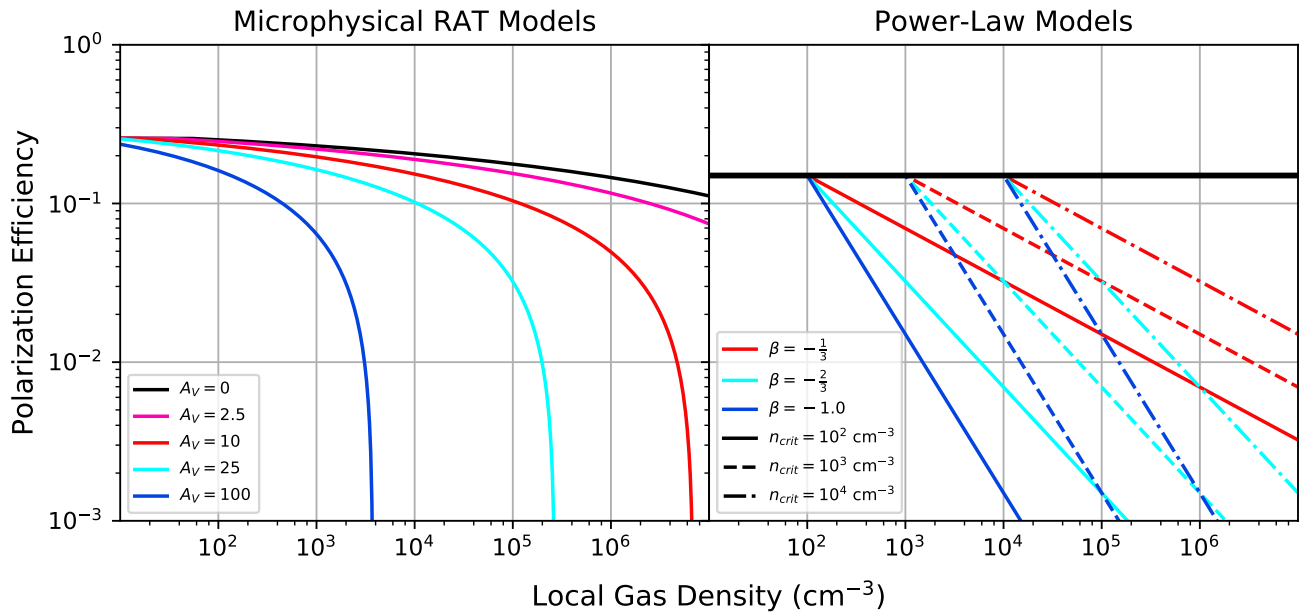}
\vspace{-8mm}
\caption{Comparison of some realizations of the power-law model (Equation \eqref{p0powerlaw}) for polarization efficiency (right panel) to the simple radiative torque alignment prediction discussed in Section \ref{section:PLDM} for several constant extinction choices (left panel). The RAT prediction was produced by combining Equations \eqref{CLalg} and \eqref{p0RAT} and assuming a constant $A_V$ as a parameter. The $A_V$ choices (2.5, 10, 25, and 100) were chosen to exaggerate the effects of high $A_V$ on polarization efficiency. The power-law models all adopt a value for $\overline{p_0} = 0.15$.}
\label{fig:ratmodel}
\end{figure*}

\begin{figure}
\centering
\includegraphics[width=\columnwidth]{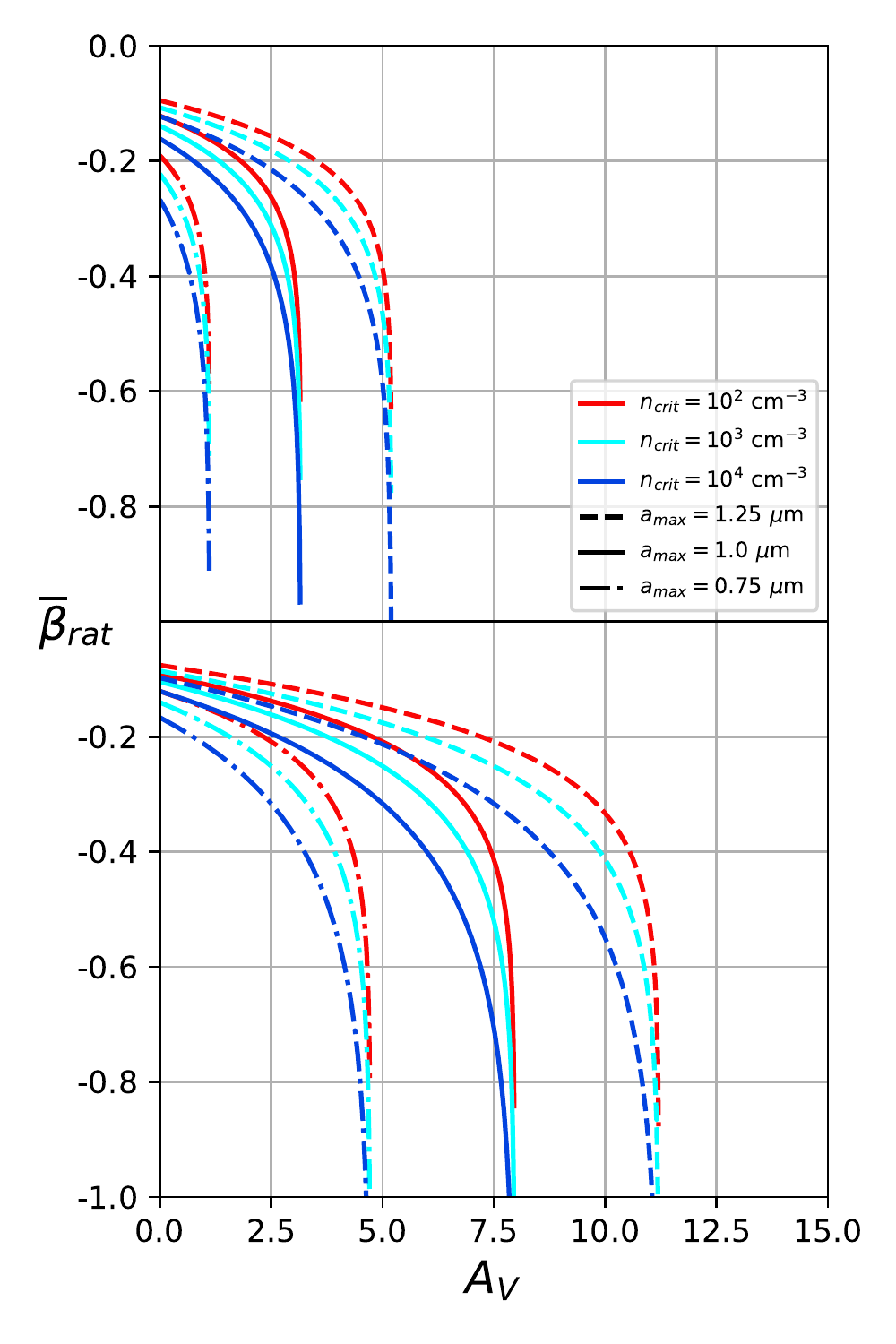}
\vspace{-8mm}
\caption{Analytic average $\overline{\beta}_{RAT}$ from Equation \eqref{betarat}, computed for two choices of maximum local gas density $n_{max}$ (top panel, $n_{max} = 10^7$ cm$^{-3}$; bottom panel, $n_{max} = 10^6$ cm$^{-3}$); three choices for critical density $n_{crit}$ (red lines, $n_{crit} = 10^2$ cm$^{-3}$; cyan lines, $n_{crit} = 10^3$ cm$^{-3}$; and blue lines, $n_{crit} = 10^4$ cm$^{-3}$); and three choices for maximum grain size $a_{max}$ (dashed lines, $a_{max} = 1.25$ $\mu$m; solid lines, $a_{max} = 1.0$ $\mu$m; and dash-dot lines, $a_{max} = 0.75$ $\mu$m); all as a function of average extinction $A_V$.}
\label{fig:ratbeta}
\end{figure}

\begin{figure*}
\centering
\includegraphics[width=\textwidth]{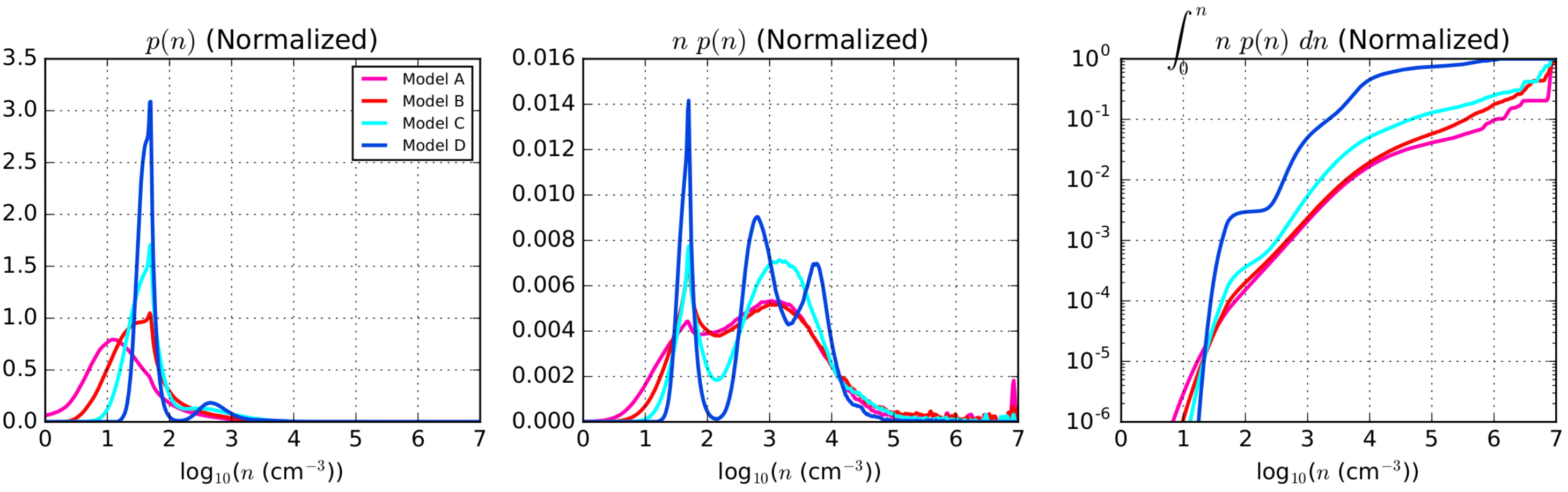}
\vspace{-6mm}
\caption{The probability density function for density, $p(n)$ (left panel); the quantity $n~p(n)$ (middle panel); and the cumulative integral $\int_0^n n~p(n)~dn$ (right panel); for all four simulations as functions of density on a logarithmic scale. The density PDFs are calculated using kernel density estimation using 50\% of the voxels of each simulation. The quantity $n~p(n)$ is a rough measure of how much of the total emission would be expected from voxels of that density, and also highlights features of the shock not visible in only the PDF, such as the secondary subshock in Model D discussed in \citet{CLKF}. The middle panel makes clear that even considering the outsize contributions of the densest voxels, their paucity reduces their contribution to the total emission. The extremely low values of the cumulative integral in the right panel for the lowest density indicate the relative unimportance of the preshock region despite dominating the overall PDF in likelihood.}
\label{fig:densitypdfs}
\end{figure*}

It is well-established observational fact that observations toward higher column density regions tend to be depolarized relative to their lower column density neighbours \citep{ALV}, and this trend has emerged as a key observational constraint on models of grain alignment and their connection to underlying magnetic structure. Quantitatively, column density and polarization fraction have significant anticorrelation in log-space \citep{PFG,PXIX,FBP,KFCL}, which suggests a power-law type dependence of decreasing polarization fraction with increasing column density, though this relationship is subject to significant scatter. Morphologically, their joint correlations have not strongly suggested a more complex relationship, being often found to be joint lognormal, though it is possible that higher resolution observations could reveal higher order structure in the future. 

The examination of the joint correlations of column density and polarization fraction for simulated clouds in Paper I revealed that underlying magnetic structure does indeed modestly influence these correlations, which are significantly affected by the apparent magnetic structure to the observer. Moreover, by choosing the appropriate level of turbulence and inclination of the line-of-sight with respect to the mean magnetic field in the synthetic observations, the polarization fraction and dispersion in polarization angles seen in the BLASTPol observations of Vela C \citep{FBP} could be very well-matched by the simulated clouds, even in joint correlation. However, pure magnetic structure effects were not capable of fully accounting for the column density-polarization fraction joint correlation, as the synthetic observations failed to match the steepness of correlation typically found in MCs. As a result, homogeneous alignment models were not sufficient to fully account for these correlations. 

It is important to assess what aspects of the modelling might give rise to this mismatch. On the one hand, it is possible that the numerical simulation models adopted in Paper I possess structures too distinct from those found in Vela C, which could account for the failures arising in the column density correlations, as column density is the most direct observation of gas structure. However, this objection may be countered by noting that the steep column density-polarization fraction correlation has been observed in many clouds, not just Vela C; presumably, the comparisons made in Paper I could be repeated with other clouds whose gas structures are significantly different from those in Vela C. Moreover, the models adopted in that work were based on plausible cloud formation scenarios \citep{IF1,I2}, and while idealized, it is unclear what characteristics other simulation paradigms (such as the turbulent periodic box) might alternatively select that would give rise to a correct anticorrelation. This situation may be clarified in future studies that might use a wider range of turbulent flow geometries in the simulations. Additionally, studies which invoke more sophisticated treatments of magnetized turbulence in the dense ISM, such as \citet{SILCC}, could provide an important check on the validity of standard turbulence treatments and simulation geometries. \\

However, it is known from grain alignment modelling (e.g. \citet{CL1}) that homogeneous alignment is not strictly accurate. Proceeding from the synthetic Stokes parameters in Section \ref{section:hetalign}, we will consider a simple form for the polarization efficiency that is agnostic with respect to particular grain alignment microphysics, in the form of a density power law:

\begin{equation}
p_0 = \begin{cases}
		\overline{p_0}~, \ ~~~~~~~~~~~~~~~~ \ n < n_{crit}, \\
        \overline{p_0} \left(\displaystyle\frac{n}{n_{crit}}\right)^{\beta}, \ ~~~ \ n \geq n_{crit}. \\
      \end{cases}
\label{p0powerlaw}
\end{equation}

\noindent Here, $n_{crit}$ is the density below which grain alignment is assumed to be unmodified, and $\beta$ is some power law index, taken to be negative. Such a model has the virtue of simplicity, making it straightforward to test the generic effects of heterogeneous alignment without demanding that the model conform to any particular model of polarization efficiency. Such a model is also dramatically simpler to implement in efficient polarized radiative transfer tools that use Equations \eqref{I}, \eqref{Q}, and \eqref{U} as their basis. For our purposes, we will take $\overline{p_0} = 0.15$ for consistency with Paper I. Quantitative estimates of this parameter may be found in \citet{PXX} and \citet{FBP}. 

A density power-law may be justified on the naive grounds that polarization is observed to vary with the projected (column) density, and therefore a local dependence on gas density may be preserved after projection. More informed justification may be offered on the grounds that the current most-accepted grain alignment theory, radiative torque alignment \citep{ALV}, is one where the local polarization efficiency is determined by grain interactions with the local radiation field. The local radiation field may be expected to vary primarily with local gas density rather than other cloud properties (such as the velocity or magnetic fields), at least to a first order approximation (i.e. neglecting the effect these fields have on turbulence and larger-scale flows in the cloud.) Further, the specific grain properties in the local region in question strongly affect the efficiency of radiative torques, and these properties are less likely to vary with the local velocity or magnetic fields. 

\subsection{Radiative Torque Alignment and Microphysical Depolarization Models} \label{subsection:RATDM}

On the other hand, a more quantitative treatment is appropriate to assuage concerns over choosing a model more convenient than accurate. Following previous analysis \citep{FP1,CL1,PJP,BCLK,PJP2}, one can obtain an expression for $p_0$ in terms of grain properties, averaged over an ensemble of $j$ grain populations, each of which has $c_j$ fractional abundance:

\begin{equation}
p_0 = \frac{\displaystyle\sum_j c_j C_{pol,j} R_j F_j}{\displaystyle\sum_j c_j C_{ran,j}~~~~~~}
\label{p0pop}
\end{equation}

\noindent ($j$ subscripts indicate the value for the $j$-th grain population.) Here, $C_{pol}$ and $C_{ran}$ are the cross sections for polarized and randomly aligned grains, respectively; $F$ is the polarization reduction due to the turbulent component of the magnetic field at small scales, as described in \citet{FP1}; and $R$ is the Rayleigh reduction factor, which is the polarization reduction due to imperfect grain alignment with the magnetic field. $F$ is typically assumed to be 1 (e.g. \citet{PJP}), equivalent to claiming that the magnetic field is sufficiently resolved in the simulations to neglect magnetic field variations within a simulation voxel. 

We can proceed from Equation \eqref{p0pop} by adopting suitable grain populations and restricting the aligned grains according to some predictions made by radiative torque alignment theory. \citet{MRN} proposed the widely used grain population model in which the number density $n_g$ of grains is proportional to a power of the size of the grains $a$:

\begin{equation}
n_g(a) \propto a^{-3.5}.
\label{MRN}
\end{equation}

\noindent Following \citet{CL1} and \citet{BCLK}, we can then assume that radiative torques result in no alignment ($R = 0$) when the grain size is less than, and total alignment ($R = 1$) provided that the grain is larger than, a minimum aligned grain size. An empirical estimate was given in \cite{CL1}:

\begin{equation}
a_{alg} = \frac{(\log_{10} n)^3 (A_V + 5)}{2800}\text{$\mu$m}.
\label{CLalg}
\end{equation}

\noindent It should be noted that this estimate depends on several simplifying assumptions, and may not be valid in all conditions or MCs. Nevertheless it is useful in extracting a meaningful interpretation for the power-law prescription in Equation \eqref{p0powerlaw}.

Submillimeter observations fall in the long wavelength limit (i.e. the dust grain sizes $a \lesssim$ $\mu$m, so $a \ll \lambda$, the millimeter/submillimeter observation wavelengths) so that both the polarized and unpolarized cross sections are proportional to the grain volume\footnote{In Equation \eqref{crosssection} we have neglected normalizing constants that contain dimensioned quantities that would yield a properly dimensioned cross section. These quantities include fundamental constants and compositional properties, which are subsumed into the polarized and unpolarized cross sections relative to the geometric cross section.} \citep{D}:

\begin{equation}
C_j(a) \propto \frac{a^3}{\lambda^2}.
\label{crosssection}
\end{equation}

\noindent Adopting a two grain (silicate and carbonaceous) population with the cross sections chosen in \citet{CL1}, and a constant gas-to-dust ratio, the polarization efficiency can be integrated over the grain populations to yield an expression in terms of $a_{alg}$ and the limits of the prescribed grain population $a_{min}$ and $a_{max}$:

\begin{equation}
p_{0,RAT} = 0.2577 \left(\frac{\sqrt{a_{max}} - \sqrt{a_{alg}}}{\sqrt{a_{max}} - \sqrt{a_{min}}}\right),
\label{p0RAT}
\end{equation}

\noindent where the number in front of the expression in Equation \eqref{p0RAT} is determined only by the relative grain population abundances and polarization cross sections (relative to geometric.) We adopt the same abundances and polarization cross sections as \citet{BCLK}, which are derived from \citet{LD1}.

To obtain a complete estimate requires closing Equation \eqref{CLalg} to obtain the extinction as a function of density, which is difficult in most circumstances. Nevertheless, if we adopt a form for this closure, one may obtain a complete heterogeneous alignment prescription in terms of the local gas density. A simplistic, though natural, choice is to assume a constant (average) extinction for the whole cloud. This choice ignores the expected significant variations each voxel might see in terms of an anisotropic radiation field, and so must be understood to be a crude estimate. This changes $A_V$ into a model parameter which is independent of the local gas density. Under these assumptions, we produce a prediction of the grain alignment efficiency that is shown in Figure \ref{fig:ratmodel}, under the assumption that $a_{min} = 0.005$ $\mu$m and $a_{max} = 1.0$ $\mu$m. We used a wide range of $A_V$ in these predictions to exaggerate the effect of polarization efficiency at extremely high extinction; $p_0\sim 0.1$ for even very high extinction ($A_V \sim 25$ to $100$) and high density ($n \sim 10^3$ to $10^4$ cm$^{-3}$) regions, which is consistent with the previous results that even deeply embedded grains can remain aligned \citep{CL1}. 

\subsection{Approximating Radiative Torque Depolarization} \label{subsection:approx}

It bears considering what part of these polarization efficiency predictions to approximate, as these models are obviously not purely power-laws and different choices in Equation \eqref{p0powerlaw} could be made to approximate these models, provided they are taken at face value. Given Equation \eqref{p0RAT}, we can compute the appropriate power-law index at a given density by evaluating the logarithmic derivative of Equation \eqref{p0RAT}:

\begin{equation}
\beta_{RAT} = \frac{d \log_{10} p_0}{d \log_{10} n~} = -\frac{\frac{3}{2}\Lambda_0^{3/2}\left(\log_{10} n\right)^{1/2}}{1 - \Lambda_0^{3/2}(\log_{10} n)^{3/2}}.
\label{dlogpdlogn}
\end{equation} 

\noindent Here, we have defined the scaling 

\begin{equation}
\Lambda = \left(\frac{A_V + 5.0}{4800.0}\right)^{1/3}\left(\frac{a_{max}}{\mu\text{m}}\right)^{-1/3} \log_{10} n = \Lambda_0 \log_{10} n,
\label{scale}
\end{equation}

\noindent We can then compute the average $\beta_{RAT}$ over $\Lambda$ to estimate an appropriate choice for our power-law index:

\begin{equation}
\overline{\beta}_{RAT} = \frac{-\frac{3}{2} \Lambda_0}{\Lambda_{max} - \Lambda_{min}} \int_{\Lambda_{min}}^{\Lambda_{max}} \frac{\Lambda^{1/2}}{1 - \Lambda^{3/2}} d\Lambda.
\label{betaratdef}
\end{equation}

\noindent A singularity exists at $\Lambda = 1$, so care must be taken to ensure that $\Lambda_{max} < 1$.\footnote{Provided that $\Lambda_{max} > 1$, it is possible to define this integral using the Cauchy principal value. However, this singularity arises when the slope approaches negative infinity, and corresponds exactly to when $a_{alg} = a_{max}$, so defining a slope beyond this is not physical.} $\Lambda_{min}$ has a natural interpretation in the form of $n_{crit}$ provided that it the polarization efficiency has not dropped far from $\overline{p}_0$:

\begin{equation}
\Lambda_{min} = \Lambda_0 \log_{10} n_{crit}.
\label{lambdamindef}
\end{equation}

\noindent Altogether we have a single expression for the average $\overline{\beta}_{RAT}$, assuming that $\Lambda_{max} = \Lambda_0 \log_{10} n_{max} < 1$ is satisfied:

\begin{equation}
\overline{\beta}_{RAT} = \frac{1}{\log_{10}\left(\frac{n_{max}}{n_{crit}}\right)} \log_{10}\left(\frac{1 - \Lambda_0^{3/2}\left(\log_{10} n_{max}\right)^{3/2}}{1 - \Lambda_0^{3/2}\left(\log_{10} n_{crit}\right)^{3/2}} \right).
\label{betarat}
\end{equation}

\noindent This expression is plotted, as a function of $A_V$ (which controls $\Lambda$ in Equation \eqref{scale}), in Figure \ref{fig:ratbeta}, for two choices of $n_{max}$, three choices of $n_{crit}$, and three choices of $a_{max}$, to illustrate the dependence of $\overline{\beta}_{RAT}$. Generally, choosing a more negative $\beta$ corresponds to adopting a larger average $A_V$, though the precise value is determined by choices in $n_{max}$ and $a_{max}$; additionally, the choice of $n_{crit}$ matters as a secondary effect. It should be noted that this analysis depends critically on the empirical relation, Equation \eqref{CLalg}, and its specific functional dependence with gas density as reported in \citet{CL1}. Informed by the RAT theory, it is sensible that the gas density should enter in explicitly, as it affects the grain dynamics directly through drag, but as Equation \eqref{CLalg} is empirically determined and based on specific choices for cloud properties and radiation field anisotropy, further modelling efforts may make this choice subject to revision. Nevertheless, we have established that $\beta$ is a parameter linked to grain alignment whose value is tied to both alignment microphysics and local conditions.

\subsection{Justifying Choices for $\beta$ and $n_{crit}$} \label{subsection:justify}

To determine which density regions contribute the most to dust emission, we computed the density PDF of voxels in our simulations using kernel density estimation (KDE), shown in the leftmost panel of Figure \ref{fig:densitypdfs}. This was then used to compute the quantity $n~p(n)$ and the cumulative integral $\int_0^n ~n~p(n)~dn$, displayed in the middle and rightmost panel of Figure \ref{fig:densitypdfs} respectively. While the differences between the simulations are apparent in the density PDFs (the lowest turbulence Model D shows the most distinct structures including a well separated pre-shock and post-shock region, whereas the highest turbulence Model A has only a single well-defined peak), the quantity $n~p(n)$ reveals even more information; consider the distinct second peak in the high density region of the Model D PDF, which indicates the dense secondary shock discussed in \citet{CLKF}. Indeed, this oblique magnetized shock structure provides a simple if crude means to choose three representative choices for $n_{crit}$: $n_{crit} = 10^4$ cm$^{-3}$ applies the power-law depolarization only the densest voxels, which tend to arise due to self-gravity rather than the larger scale secondary shock; $n_{crit} = 10^3$ cm$^{-3}$ applies the power-law depolarization mostly to voxels in the secondary shock; and $n_{crit} = 10^2$ cm$^{-3}$ applies the power-law depolarization to most of the entire post-shock region. This rule of thumb is moderately weakened considering that the dense secondary shock isn't perceptible in the Model A, B, and C, but nevertheless it retains some value in guiding understanding of the role of $n_{crit}$. 

Considered as a population, each voxel should contribute to the integrated total intensity roughly $n~p(n)$, since the emission is density-weighted (see e.g. Equation \eqref{I}). Even when considering the outsize contribution of the densest voxels, the middle panel demonstrates that their contributions are rapidly reduced due to their extreme paucity. Additionally, by considering the cumulative integral, we see that the contributions from the lowest density voxels, including the preshock region, are nearly negligible, below a part in a thousand for Models A, B, and C, and perhaps a few parts in a thousand for Model D. Evidently, the critical region to model is the intermediate density regime, and suggests that the Stokes parameters are most strongly affected by voxels within that density range.
 
In the interest of accuracy, we could have adopted a more complex form, such as Equation \eqref{p0RAT} instead of Equation \eqref{p0powerlaw}, in order to better approximate the radiative torque models. However, by adopting this simple power-law form for the polarization efficiency, we can more easily test the general effects on the synthetic polarization observables that heterogeneous alignment can have, without relying too heavily on the complicated dynamics of microphysical grain alignment, which are strongly dependent on exact local conditions, i.e. the local radiation field and local grain populations, as discussed in \citet{CL1,PJP,BCLK} and \citet{PJP2}. Additionally, by restricting the parameters choices it becomes simpler to understand these effects in an unambiguous way. Nevertheless, we plan to explore using more sophisticated models in future work. 

\section{Synthetic Observations with Power-Law Depolarization Models} \label{section:plresults}

\subsection{Distributions of Polarization Fraction and Dispersion in Polarization Angles} \label{section:pl1D}

\begin{figure*}
\centering
\includegraphics[width=0.49\textwidth]{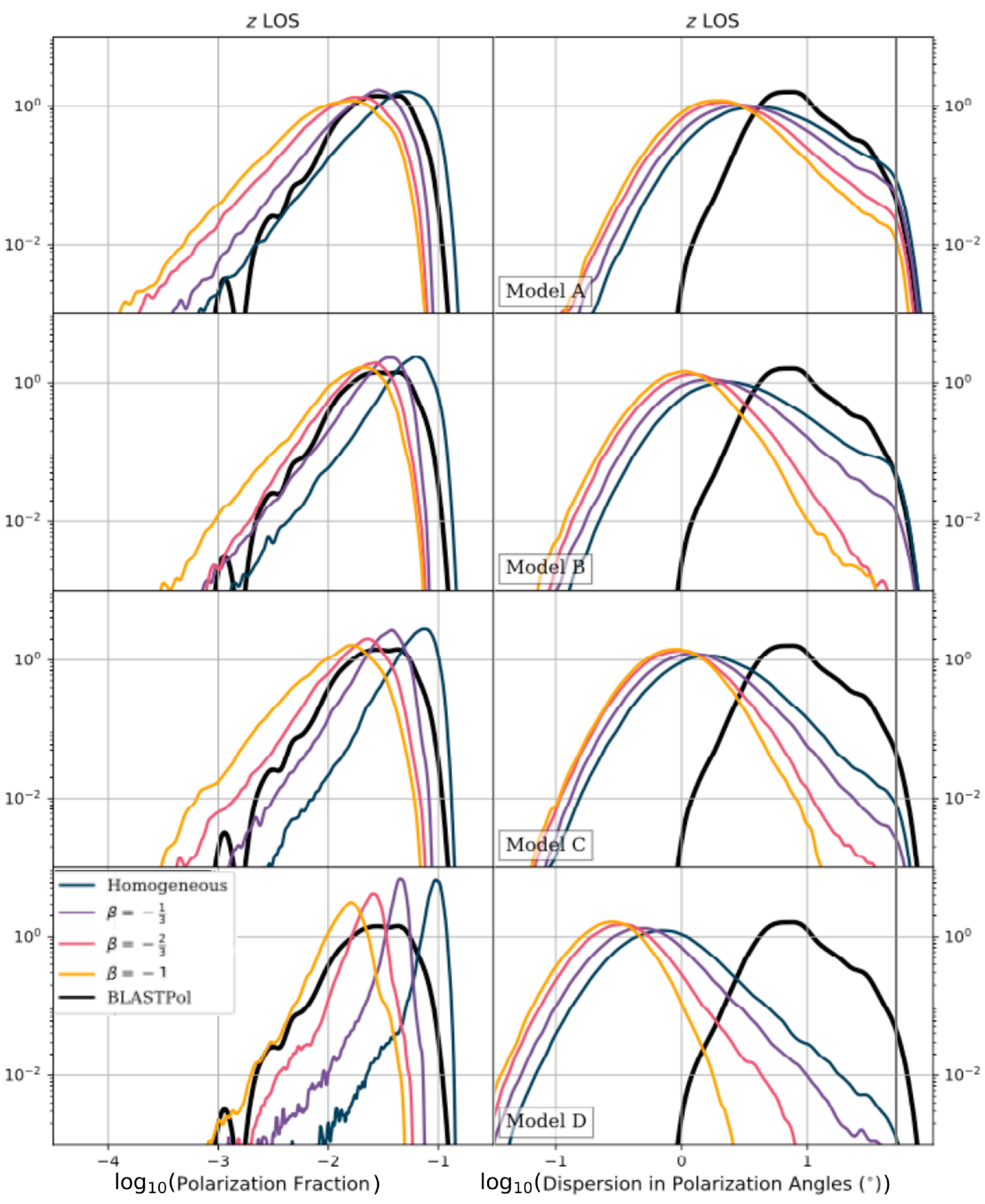}
\includegraphics[width=0.496\textwidth]{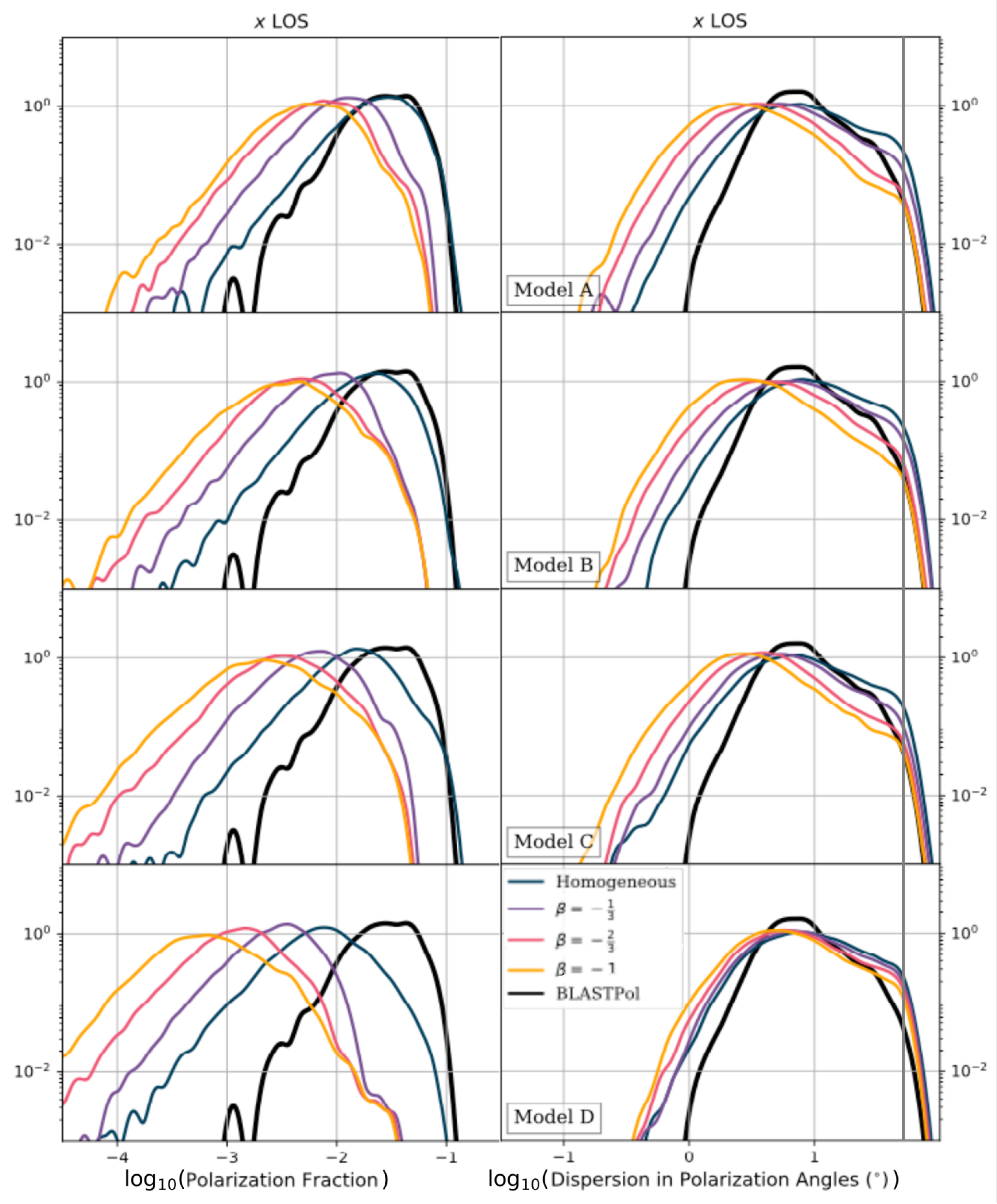}
\caption{One dimensional PDFs of the polarization fraction and dispersion in polarization angles for Models A (top row), B (second rows from the top), C (second rows from the bottom), and D (bottom rows). The left two columns contain the synthetic observations from the $z$ line-of-sight, and those in the right two columns contain those from the $x$ line-of-sight. Each plot contains the PDFs obtained under homogeneous alignment with $p_0 = 0.15$ (dark blue) as well as power-law depolarization models (Equation \eqref{p0powerlaw}) with $\beta = -1/3$ (purple), $-2/3$ (dark pink), and $-1$ (yellow); $n_{crit} = 10^2$ cm$^{-3}$ for each of these synthetic observations. The equivalent PDF of the BLASTPol observations of Vela C are annotated on each plot in black for comparison. The characteristic value, $\mathcal{S} = \pi/\sqrt{12}$, is noted as a vertical grey line on the $\mathcal{S}$ distributions.}
\label{fig:1DPL}
\end{figure*}

\begin{figure}
\centering
\includegraphics[width=\columnwidth]{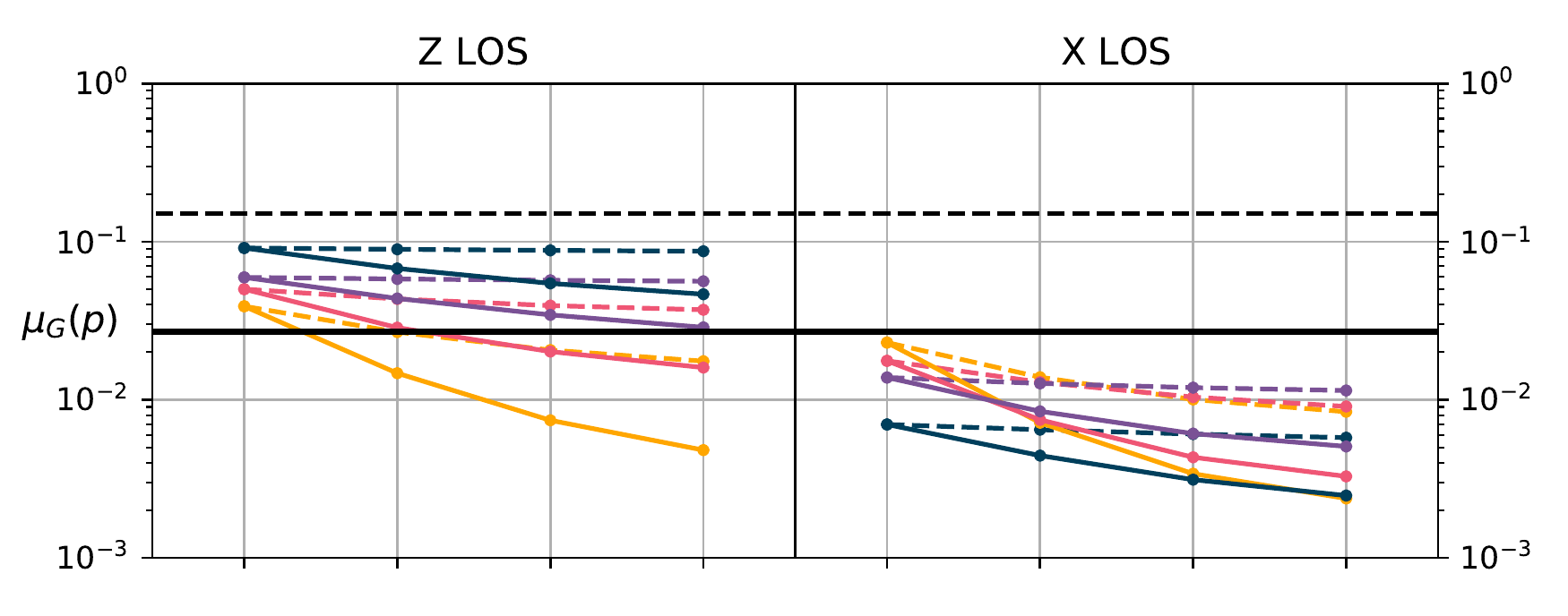}\\
\includegraphics[width=\columnwidth]{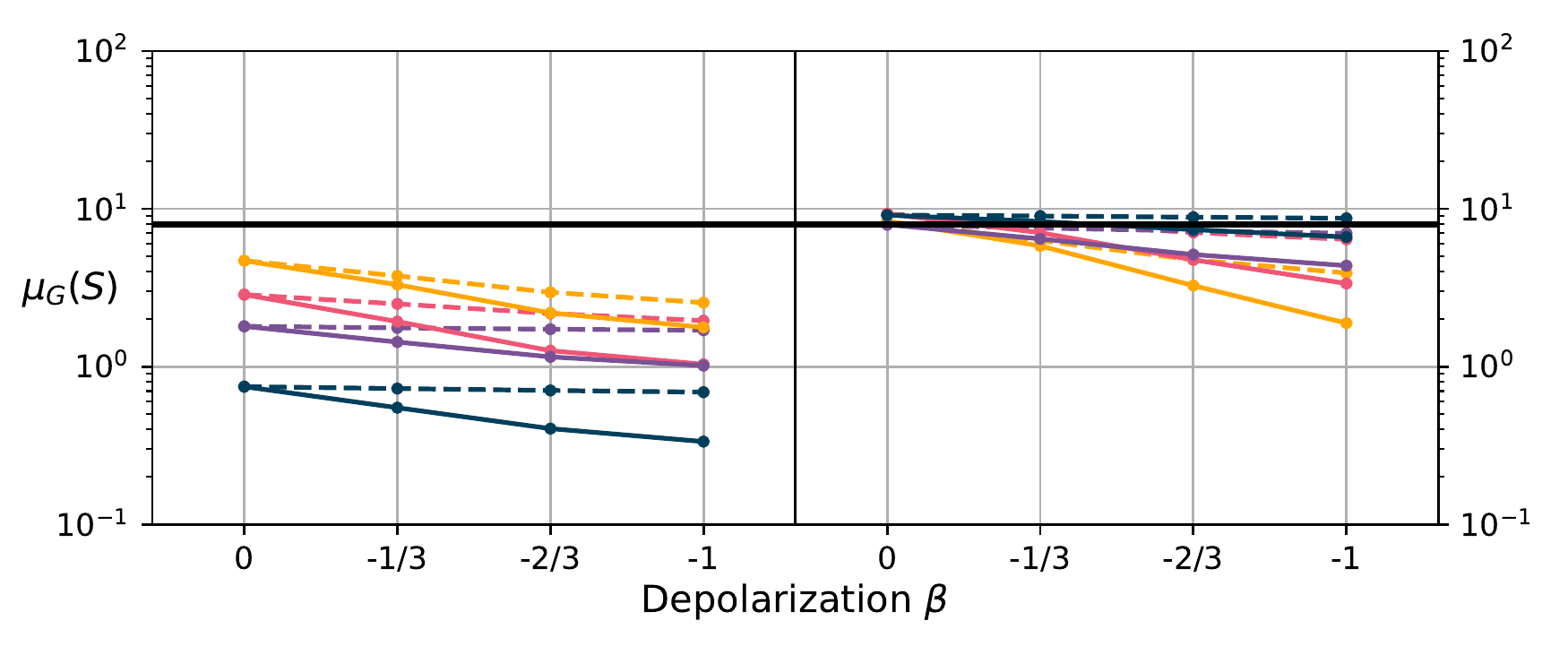}
\vspace{-8mm}
\caption{Geometric means of the polarization fraction (top row) and dispersion in polarization angles (bottom row) as a function of the power-law depolarization model $\beta$, shown in linear scale. Solid lines indicate $n_{crit} = 10^2$ cm$^{-3}$, and dashed lines indicate $n_{crit} = 10^3$ cm$^{-3}$; colours indicate the simulation, with Model A in yellow, B in dark pink, C in purple, and D in dark blue. The adopted $\overline{p_0} = 0.15$ is shown (black dashed line) in the plot of polarization fraction geometric mean; the BLASTPol derived geometric means are shown in solid black. $\beta = 0$ corresponds to a homogeneous alignment model.}
\label{fig:means}
\end{figure}

\begin{figure}
\centering 
\includegraphics[width=\columnwidth]{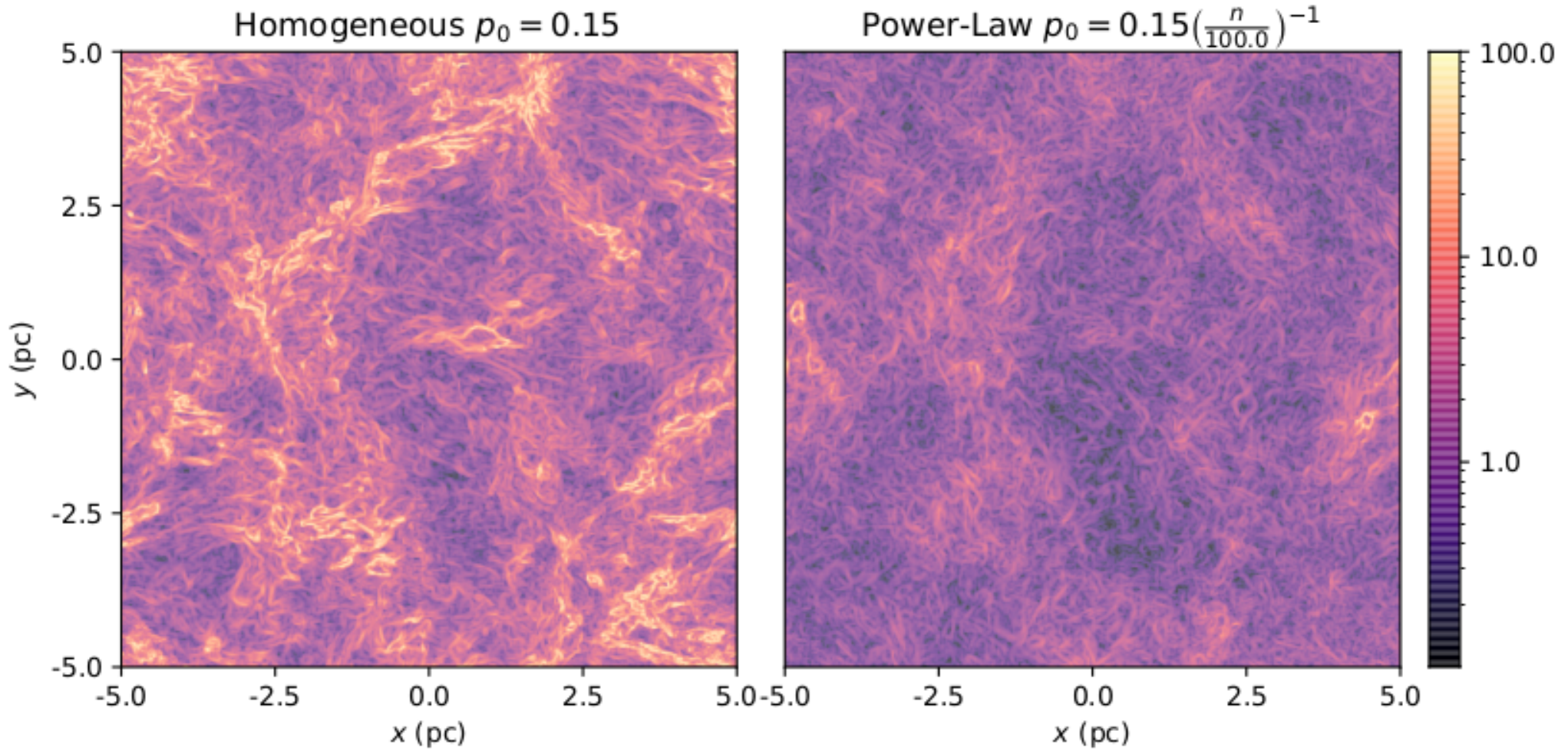}
\vspace{-6mm}
\caption{Comparison of two images of the dispersion in polarization angles $S$, computed for two choices in depolarization model for Model B viewed from the $z$ line-of-sight. The left image indicates a homogeneous alignment model, and the right image a power-law depolarization model with $n_{crit} = 10^2$ cm$^{-3}$ and $\beta = -1$. Note that much of the filamentary high-$\mathcal{S}$ structures are significantly reduced in the right image, in some cases vanishing entirely, yet much of the lower $\mathcal{S}$ structures remain.}
\label{fig:Scompare}
\end{figure}

Modifications to the intrinsic polarization efficiency, $p_0$, might be expected to have a larger effect on the polarization fraction than the polarization angle. Given that the polarization fraction (Equation \eqref{p}) is roughly linear in Stokes $Q$ and $U$, whereas the polarization angle (Equation \eqref{ch}) is computed from the ratio of the Stokes $Q$ and $U$, modifications to $p_0$ should present themselves roughly to first order in polarization fraction $p$ but only at higher orders in the polarization angle $\chi$. It follows that the dispersion in polarization angles $\mathcal{S}$ could be similarly less sensitive to heterogeneous alignment, at least in the crudest approximation; indeed, under homogeneous alignment, the dispersion in polarization angles is totally unaffected by the choice of $p_0$. This expectation can be tested by considering how the PDFs of $p$ and $\mathcal{S}$ change under different heterogeneous alignment models. 

As the relative orientation of the mean magnetic field to the observer's line-of-sight was found to be a controlling factor affecting the synthetic observations in Paper I and \citet{CKLFM}, we will consider two configurations: the line-of-sight in which the mean magnetic field is nearly parallel to the line-of-sight (the $x$ line-of-sight, see Figure \ref{fig:geometry}); and the line-of-sight in which the post-shock region is nearly face-on to the observer, and the mean magnetic field is nearly perpendicular to the line-of-sight and therefore mostly in the plane-of-sky (the $z$ line-of-sight.)\footnote{The $y$ line-of-sight has been omitted for brevity as it was found in Paper I that it has nearly the same properties as the $z$ line-of-sight, but poorer statistics due to its edge-on view of the post-shock region.} These PDFs for both the $x$ and $z$ lines-of-sight, for three choices of power-law index ($\beta = -1/3, -2/3, -1$) for critical density $n_{crit} = 10^2$ cm$^{-3}$, along with the unmodified homogeneous alignment PDFs for comparison, are found in Figure \ref{fig:1DPL}. The general effect on each observable can be explored by plotting how a measure of central tendency changes with each polarization model; we present the change in geometric means\footnote{The quantitative difference between the geometric means, medians, and modes (PDF peak location) were found to be very near each other, so we chose the geometric mean for definiteness.}, denoted $\mu_G$, for both lines of sight in Figure \ref{fig:means}. Interestingly, the expectation - that $p$ is sensitive to grain alignment physics, but $\mathcal{S}$ less so - exactly holds in only some cases. 

Adjustments to $\mu_G(p)$ under the effect of heterogeneous alignment models should be understood in the context that $\overline{p_0}$ in Equation \eqref{p0powerlaw} serves as an overall homogeneous or scalar component to the polarization efficiency. This is not really a free parameter, being determined by the grain populations (size distribution and composition) and potentially other effects, such as intrinsic alignment inefficiencies common to all grains in all conditions. Nevertheless this parameter remains partially unconstrained, and so the polarization fraction populations may still be scaled through a choice of this parameter to match observations, within reason (i.e. extremely small or large values are not physically sensible). Changes to $\mu_G(p)$ thus should be considered relative to the pure homogeneous alignment cases, and the relative change should be interpreted as additional depolarization due to elimination of contributions from the highest volume density regions through the application of a heterogeneous alignment model, which physically arises due to decreased polarization efficiency. One should also consider that the overall polarization level in observations must be calibrated against a suitable heterogeneous alignment model and dust grain population in order to ascertain a true $\overline{p_0}$. If one assumes homogeneous alignment to be true in an observational context, one may be led to underestimate $\overline{p_0}$ by fitting against an incomplete model of depolarization, even if one can fully account for depolarization effects due to inclination and cancellation within the line-of-sight. 

Since a lower critical density $n_{crit}$ applies the power-law depolarization to more voxels within the simulation, and applies a stronger depolarization effect to voxels with the same volume density, the effects of our power-law model are exaggerated as $n_{crit}$ is reduced. Within the $z$ line-of-sight, we find that $\mu_G(p)$ is (as expected) significantly affected by power-law depolarization models, with more turbulent simulations corresponding to a stronger effect. This effect is also present, though more muted, within the $x$ line-of-sight, which is already quite low to begin with; there appears to be no general trend with level of turbulence here. The $\mu_G(\mathcal{S})$ within the $x$ line-of-sight remains closest to the expected behaviour, in which $\mu_G(\mathcal{S})$ changes the least, especially considering Model D; the higher turbulence simulations show larger differences in $\mu_G(\mathcal{S})$ with more negative $\beta$ or smaller $n_{crit}$. Within the $z$ line-of-sight, there is less difference between the simulations in the magnitude of the response of $\mu_G(\mathcal{S})$ to power-law depolarization models, as each seems to be reduced by a comparable amount, as seen in Figure \ref{fig:means}. The different levels of $\mu_G(\mathcal{S})$ instead seem to merely reflect the previously established fact that the overall level of the dispersion in polarization angles is lower for lower turbulence simulations.

In Paper I, a probability excess feature in the $\mathcal{S}$ PDFs near the characteristic value of $\pi/\sqrt{12}$ was identified, and suggested to be due to the filamentary features in $S$ (as distinguished from filamentary structures in column density; see e.g. Figure \ref{fig:Scompare}). In examining the $\mathcal{S}$ PDF morphologies in Figure \ref{fig:1DPL}, we find a secondary effect of applying our heterogeneous alignment models: the suppression of this probability excess as progressively more negative $\beta$ or lower $n_{crit}$ is applied. We present images of $\mathcal{S}$ under homogeneous alignment and a strong power-law model to show this visibly in Figure \ref{fig:Scompare}. The filamentary $\mathcal{S}$ features are significantly suppressed under the strong power-law model, which might strengthen the association of the excess probability feature with the filamentary $\mathcal{S}$ features. This suppression may provide some clues as to the origin of these filamentary $\mathcal{S}$ features: by suppressing the contributions from the highest density voxels the most, the power-law heterogeneous alignment prescription has revealed that the highest $\mathcal{S}$ filamentary features could arise from emission contributions from the intermediate to higher density regions within the line-of-sight, at least from the $z$ line-of-sight perspective, wherein the ordering influence of the mean magnetic field is apparent to the observer. On the other hand, since there is not a strong correlation between $S$ and column density (see Section \ref{section:plCD} below), these regions probably do not coincide with the self-gravitating regions; instead, they could arise at the intermediate-high density regions found within every line-of-sight, whose existence can be accounted for by referring to the dense secondary shock common to all oblique magnetized shocks described in \citet{CLKF}. However, we emphasise that separating these features from the one-dimensional PDFs (as presented in Figure \ref{fig:1DPL}) is not a simple matter, as simple thresholding is not sufficient to separate these features, so we must limit our conclusions. 


In Paper I, the importance of the level of the $\mathcal{S}$ population (i.e. $\mu_G(\mathcal{S})$) and the shape of the $p$ population (the geometric standard deviation, $\sigma_G(p)$) was emphasized, as the observable $\mathcal{S}$ cannot be adjusted through an appropriate scaling, and that narrow $p$ populations indicated high apparent magnetic order. Here, as stronger power-law depolarization models are applied, the most dramatic changes to $\sigma_G(p)$ are in the least turbulent Model D, though in general this effect is modest, and the width of the BLASTPol $p$ population remains relatively well-matched in general to the higher turbulence models; reductions to $\mu_G(p)$ from the power-law depolarization models can be simply accounted for by adjusting $\overline{p_0}$. The $\mu_G(S)$ of BLASTPol also remains best matched by the $x$ line-of-sight $\mathcal{S}$ populations, even under strong power-law models; though it was argued that the $\mu_G(\mathcal{S})$ for the $z$ line-of-sight might be raised by high levels of turbulence, we have no direct evidence that this is possible (even with the inclusion of the higher turbulence/lower magnetization Model A in this work).

These results renew the importance of the $\mathcal{S}$ observable as a diagnostic of magnetic structure. Not only have we argued that the bulk of the $\mathcal{S}$ population is perhaps less sensitive to grain alignment microphysics than $p$ (with the exception of the high-$\mathcal{S}$ filamentary features that are suppressed under our power-law depolarization prescription), but we have also demonstrated that heterogeneous grain alignment is likely to, if anything, reduce $\mu_G(\mathcal{S})$. As a result, this intensifies the disagreement presented in Paper I: because any line-of-sight that is insufficiently parallel to the mean magnetic field tended to have lower $\mu_G(\mathcal{S})$ than presented in the BLASTPol observations of Vela C, it is even more unlikely that Vela C has a mean magnetic field mainly in the plane-of-sky. Further, though we do not consider beam convolution in this work, in Paper I it was argued that beam convolution primarily affects the shapes, not the means, of observable populations, given the observed effect of applying a beam convolution: i.e. it was found to narrow $\sigma_G$ but not affect $\mu_G$, for both $p$ and $\mathcal{S}$. These results should also be considered alongside the results presented in \citet{CKLFM}, which provided another estimate of the magnetic field inclination of Vela C relative to the observer, and found that the this inclination angle should be high. As a result, we are led to conclude that MCs with the highest overall $\mathcal{S}$ populations (i.e. highest $\mu_G(\mathcal{S})$) likely have mean magnetic fields close to the line-of-sight of the observer. 

\subsection{Joint Correlations With Column Density} \label{section:plCD}

\begin{figure*}
\centering
\vspace{-2.5mm}
\includegraphics[height=0.46\textheight, width=\textwidth]{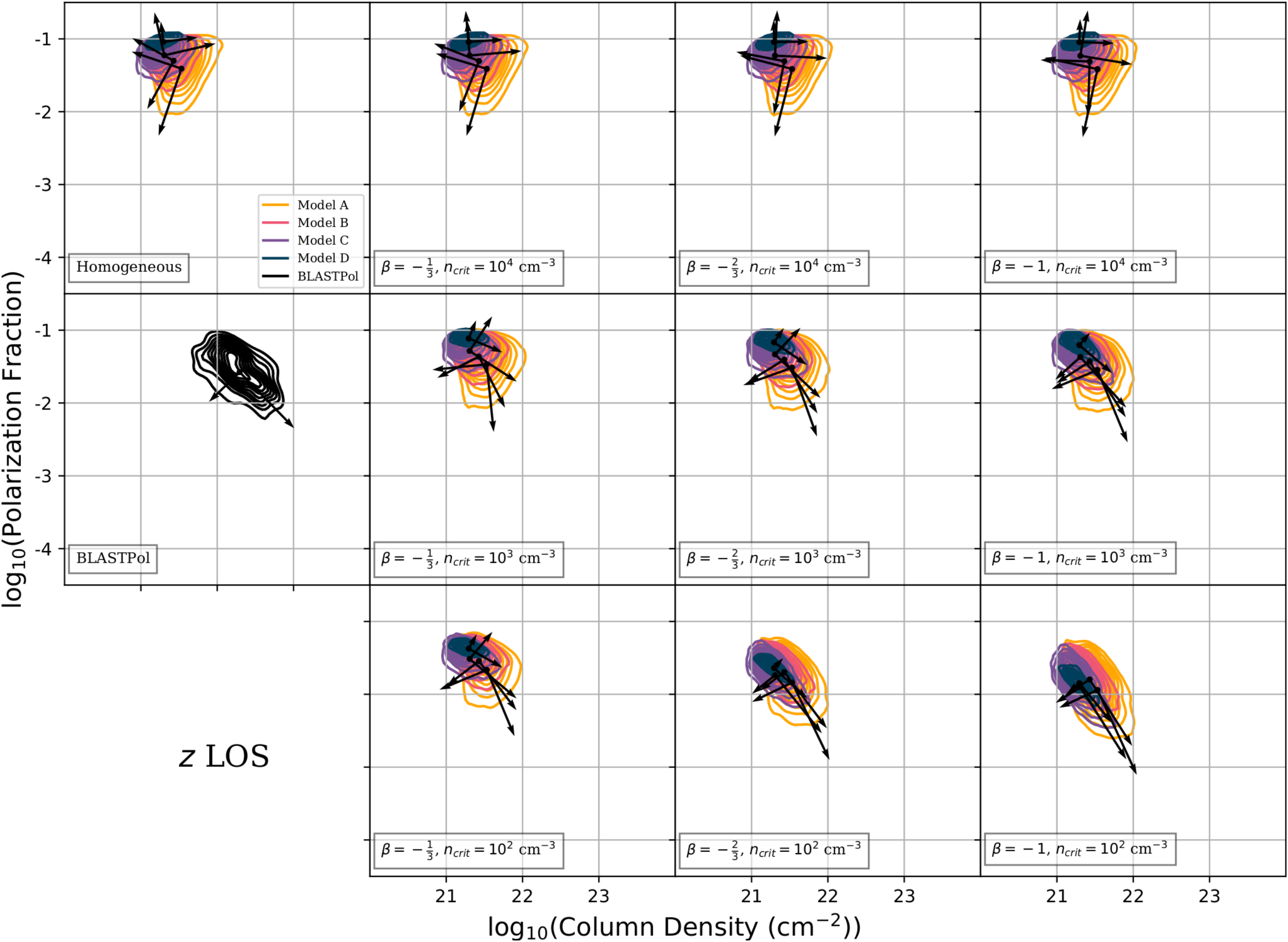}\\
\vspace{1mm}
\includegraphics[height=0.46\textheight, width=\textwidth]{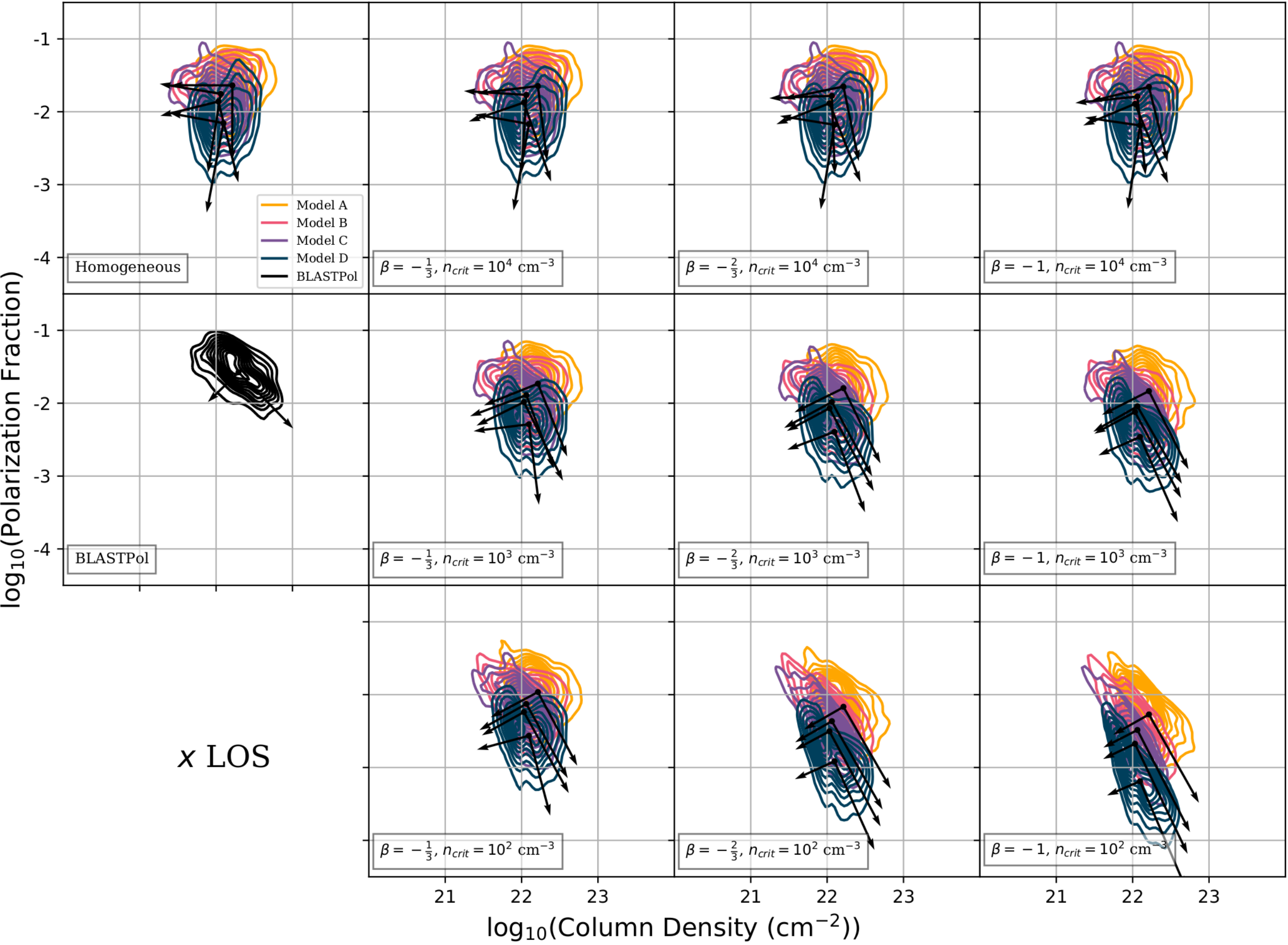} 
\vspace{-6mm}
\caption{Joint PDFs of polarization fraction with column density (contours) with annotated principal components (black vectors), whose tails are located at the coordinates of the geometric means of each observable. The top panel contains observations along the $z$ line-of-sight; the bottom, the $x$ line-of-sight. Synthetic observations under homogeneous alignment for Models A (yellow), B (dark pink), C (purple), and D (dark blue) are in the top left subpanels; in the subpanel directly below this are the equivalents for the BLASTPol observations of Vela C in black. In the remaining panels, synthetic observations using different power-law depolarization models are presented, with the second column indicating those with $\beta = -1/3$; the third column, $\beta = -2/3$; and the last column, $\beta = -1$. In those columns, those in the top row are with $n_{crit} = 10^4$ cm$^{-3}$; the middle row, $n_{crit} = 10^3$ cm$^{-3}$; and the bottom row, $n_{crit} = 10^2$ cm$^{-3}$.}
\label{fig:NpPL}
\end{figure*}

\begin{figure}
\includegraphics[width=\columnwidth]{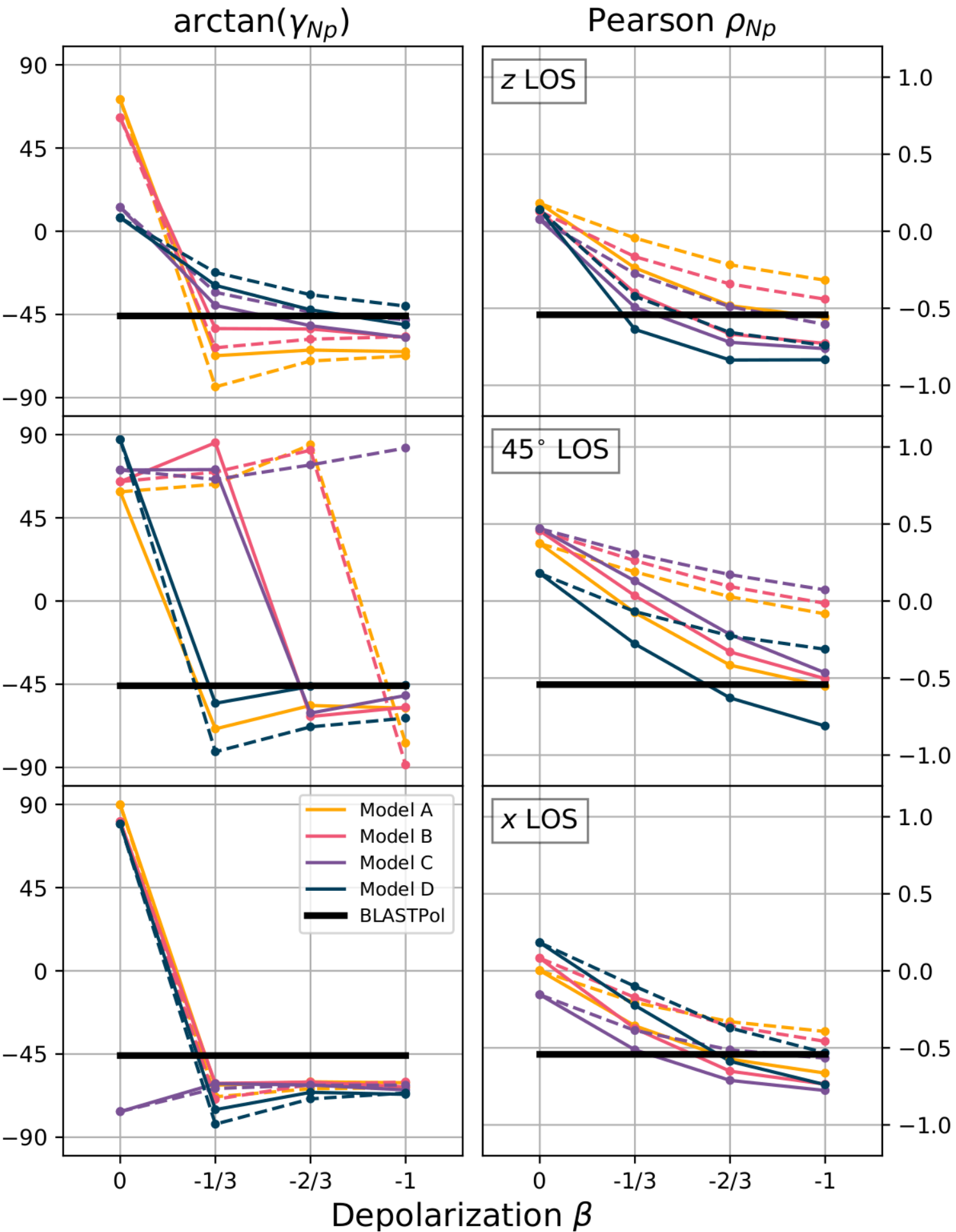}
\vspace{-4mm}
\caption{Arctangent of the PCA-implied power-law index for the column density-polarization fraction joint correlation, $\gamma_{Np}$ (left column), and Pearson correlation coefficient (in logarithmic space) between column density and polarization fraction (right column), for the $z$ line-of-sight (top row), the $x$ line-of-sight (bottom row), and a line-of-sight inclined 45$^{\circ}$ between them (middle row). Each panel contains these quantities derived from synthetic observations along these lines-of-sight for Models A (yellow), B (dark pink), C (purple), and D (dark blue), for $n_{crit} = 10^2$ cm$^{-3}$ (solid lines) and $n_{crit} = 10^3$ cm$^{-3}$ (dashed lines), with depolarization law power-law index $\beta$ on the horizontal axis. $\beta = 0$ corresponds to a homogeneous alignment model.}
\label{fig:Npstats}
\end{figure}

\begin{figure*}
\centering 
\vspace{-2mm}
\includegraphics[height=0.46\textheight, width=\textwidth]{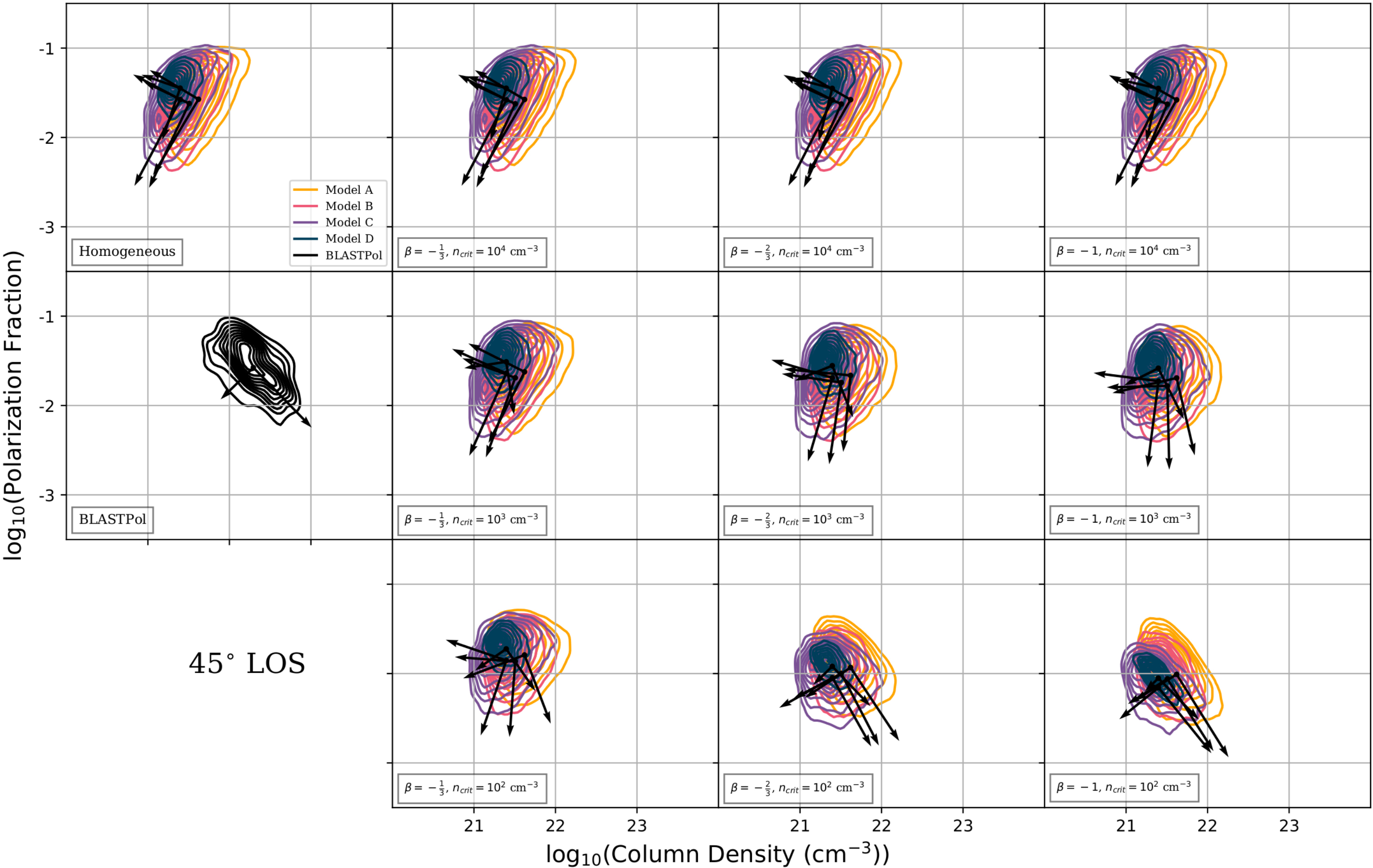}
\vspace{-5mm}
\caption{Same as Figure \ref{fig:NpPL}, but for the column density-polarization fraction correlations viewed from a line-of-sight inclined $45^{\circ}$ between the $x$ and $z$ lines-of-sight.}
\label{fig:Np45}
\end{figure*}

\begin{figure*}
\centering
\vspace{-2mm}
\includegraphics[height=0.46\textheight, width=\textwidth]{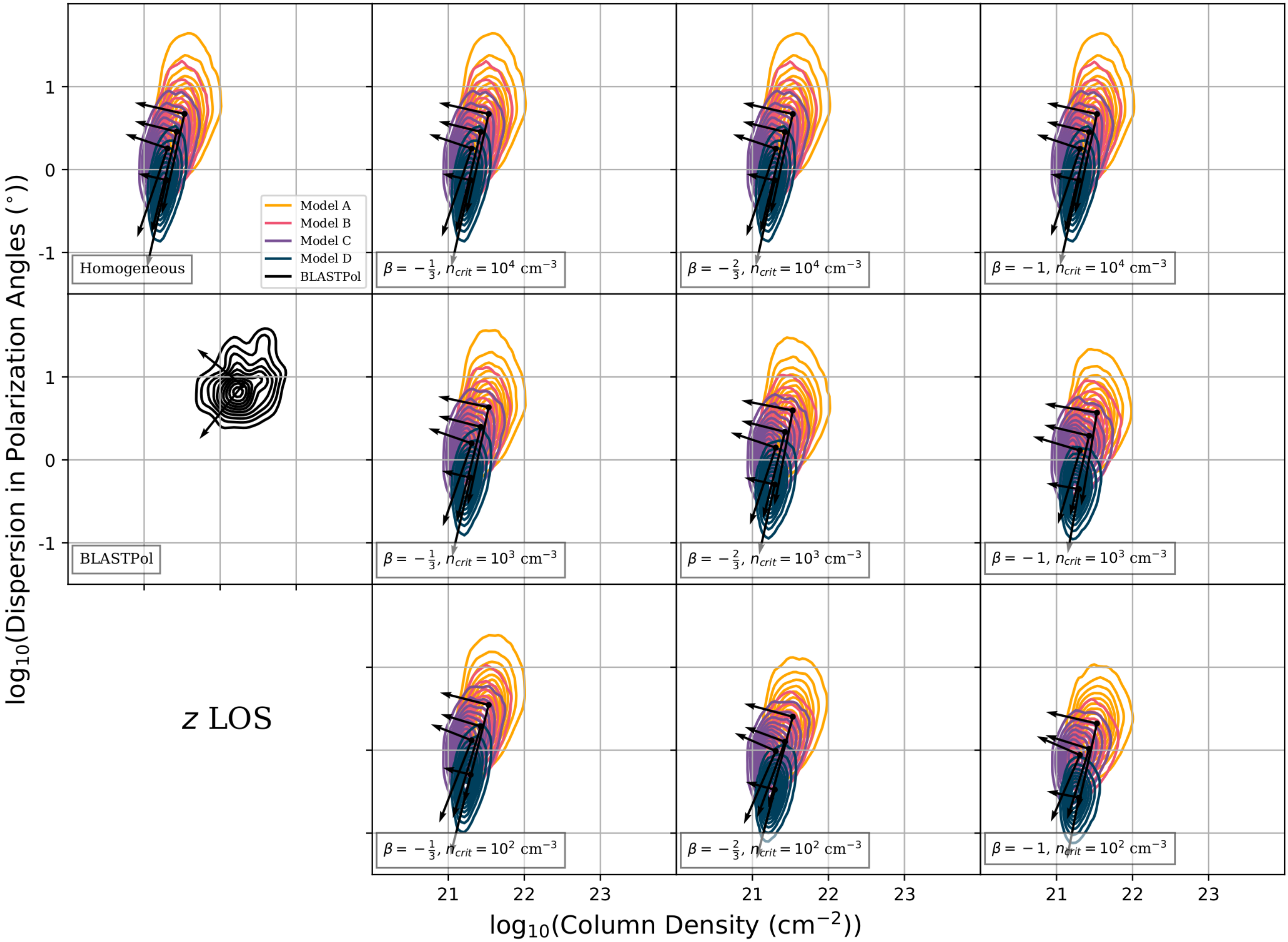}\\
\vspace{1mm}
\includegraphics[height=0.46\textheight, width=\textwidth]{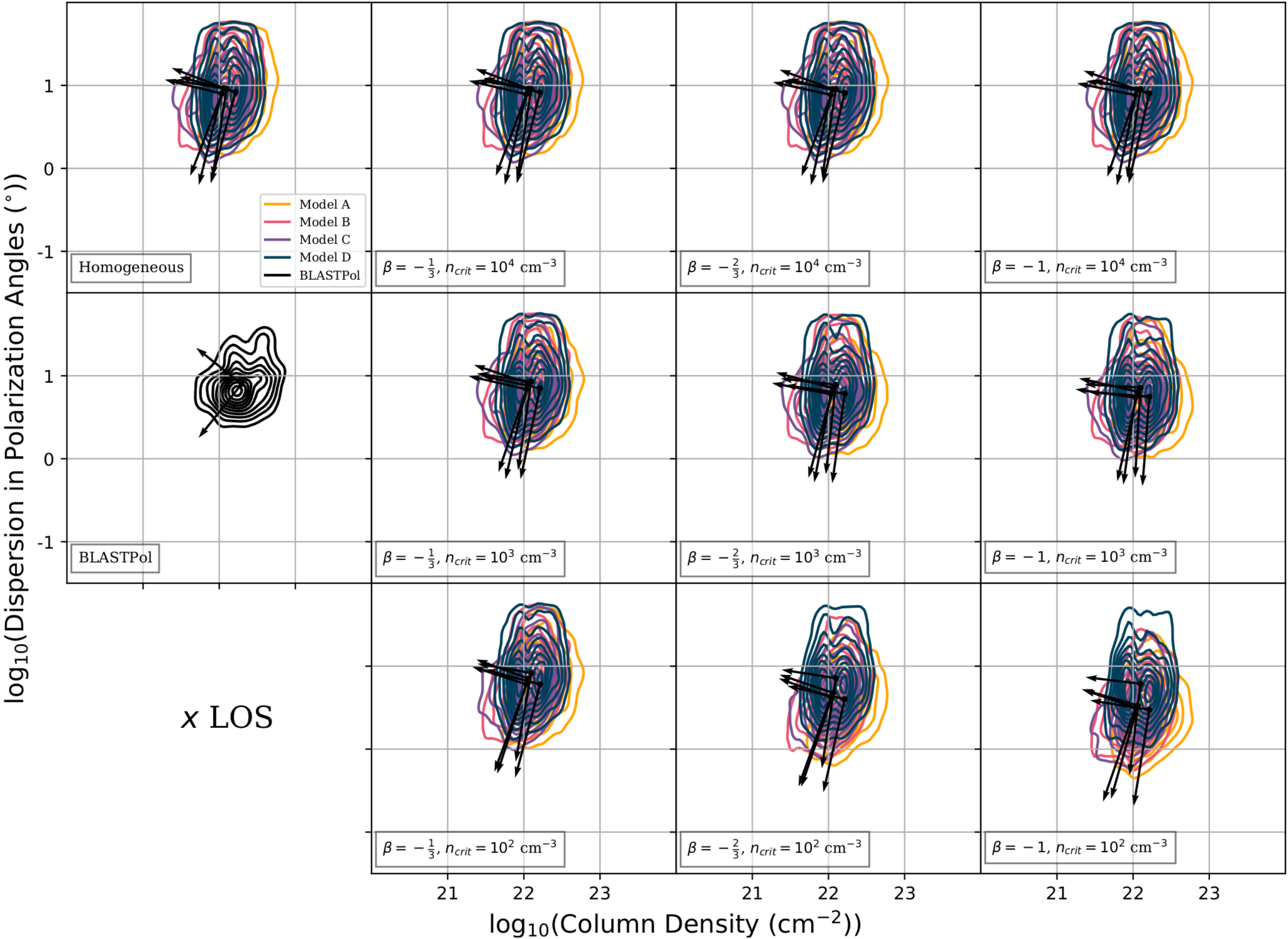}
\vspace{-5mm}
\caption{Same as Figure \ref{fig:NpPL}, but for the correlation between dispersion in polarization angles and column density.}
\label{fig:NSPL}
\end{figure*}

\begin{figure}
\centering
\includegraphics[width=\columnwidth]{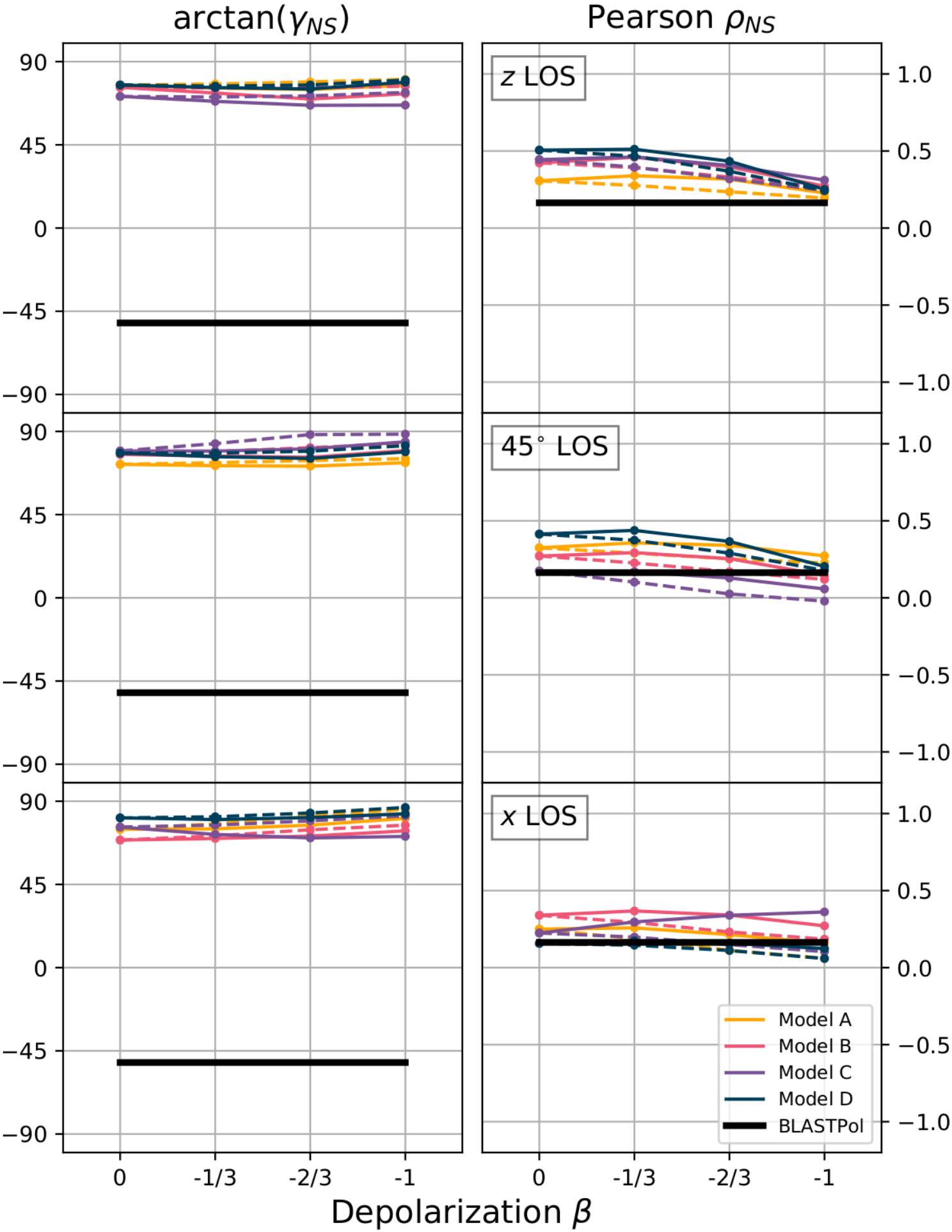}
\vspace{-4mm}
\caption{Same as Figure \ref{fig:Npstats}, but for the correlation between column density and the dispersion in polarization angles.}
\label{fig:NSstats}
\end{figure}

\subsubsection{Comparing Column Density Correlations Under Different Grain Alignment Models and Local Conditions} \label{subsubsection:coldencomp}

Being one of the chief motivations for considering heterogeneous alignment models, the correlations involving column density (particularly the column density and polarization fraction correlation) deserve careful scrutiny. As the observables change in response to adjusting the model parameters ($\beta$ and $n_{crit}$) in the power-law depolarization model, the morphology of the joint correlations can evolve in hard-to-predict ways; the principal components measure only the lowest order structure in these correlations, and can be strongly affected (i.e. rotated or dilated) by different depolarization model parameters. It is important to consider not only the principal components but also measures such as the Pearson correlation coefficient and the morphologies of the joint correlations themselves, as each piece alone provides an incomplete picture. 

To ground our subsequent discussion on heterogeneous alignment models, we first discuss our expanded set of simulations under homogeneous alignment. These models, presented in Figure \ref{fig:NpPL} (top left subpanels), reinforce the conclusions made in Paper I: neither the $x$ nor the $z$ lines-of-sight produce significant anticorrelations between column density and polarization fraction, regardless of what simulation is considered. The $z$ line-of-sight correlations tend to display a triangular shape, in which the most variance in polarization fraction is at medium column densities, tapering off at low and high column densities, with a flat top; the principal components do not suggest a simple, nearly linear relationship between these two observables, as seen in Vela C by BLASTPol \citep{FBP}. The $x$ line-of-sight joint correlations display little in the way of any higher order structure, and are generically uncorrelated. The joint distribution shape appears to be driven mainly by pure uncorrelated variance in the observables rather than by any significant joint covariance. One exception does bear mentioning: when viewed from the $x$ line-of-sight, the Model D column density-polarization fraction differs from the other simulations. Where the others are generically unimodal in a bivariate sense (suggesting a single population cluster), the Model D joint correlation suggests a bimodal distribution, principally along the column density axis. This is simply interpreted in the context of the density distribution shown in Figure \ref{fig:densitypdfs} and the secondary shock discussed in \citet{CLKF}: an edge on view of the post-shock region shows this secondary post-shock region in column density.\footnote{This can be seen in the two-peaked column density PDFs shown in Figure 7, and visually in the pixel scale edge on column density maps of Figure 2, in Paper I.} Since this dense subshock is only clearly visible in Model D, it is not surprising that this bimodality is not encountered otherwise. In either case, the polarization fraction appears generally uncorrelated with column density, in contradiction with established observational fact (see, e.g., the lower left panel of either subpanels in Figure \ref{fig:NpPL} or \citet{FBP} for the BLASTPol results, as well as \citet{PFG, ALV, PXIXalt}).

The failure to find a robust and steep (in log-space) anti-correlation under homogeneous alignment is somewhat surprising in view of the analytical work of \citet{J92} (see also \citet{J89}), which shows that the polarization fraction decreases with the column density naturally in a turbulent medium if there is a random component to the magnetic field and the random component is decorrelated over a relatively small column density interval (compared to the maximum of the region). The exact reason for this discrepancy is unclear. It is likely that in numerical simulations of turbulent magnetized clouds including self-gravity (such as ours; see also e.g. Figure 8 of \citet{FG1}, and \citep{PXX}) there is no clean separation of ordered and random components of the magnetic field and no well-defined column density interval over which the random field component becomes decorrelated.  Whether there is a way to generate the kind of turbulent magnetized clouds envisioned in \citet{J92} in MHD simulations remains to be determined.


To consider the introduction of the power-law depolarization models quantitatively, we present in Figure \ref{fig:Npstats} the arctangent of the implied power-law indices, $\gamma_{Np}$\footnote{The implied power-law index corresponds to the depolarization parameter discussed in Paper I, which is the slope of the principal component corresponding to the correlation/anticorrelation between the two observables. The arctangent of this quantity simply yields the angle, measured from horizontal, that the principal component has been rotated through. Because the correlations can be relatively weak and are in general (relative to column density) heteroskedastic in log-space, we choose the positive/negative slope principal component based on whether the Pearson coefficient is positive/negative to provide an unambiguous criterion.}, of the resulting column density-polarization fraction correlations and the Pearson correlation coefficients, for both the $x$ and $z$ lines-of-sight as well as 45$^{\circ}$ inclination between them (presented in Figure \ref{fig:Np45}). We show only those models with $\beta = -1/3$, $-2/3$, and $-1$, and $n_{crit} = 10^2$ cm$^{-3}$ and $10^3$ cm$^{-3}$, as those models with $n_{crit} = 10^4$ cm$^{-3}$ differ relatively little from the homogeneous alignment models. For both the $z$ and the $x$ lines-of-sight, consistent trends emerge in both cases: the application of power-law depolarization models tend to increase the strength of the anticorrelation (more negative Pearson $\rho_{Np}$) and rotate the principal components toward $\arctan \gamma_{Np} \sim -45^{\circ}$ (or $\gamma_{Np} \sim -1$), which is near the BLASTPol value, $\gamma_{Np} = -1.029$. The $z$ line-of-sight, as has been noted in \citet{KFCL}, displays a graded response between the different simulations, as higher turbulence/lower magnetization models are affected more strongly by our power-law depolarization models in rotating the principal components, but remain less correlated (less negative $\rho_{Np}$). On the other hand, in the case of the $x$ line-of-sight, the principal components are rotated near to the BLASTPol value, indicating a much steeper power-law. These slopes are tightly clustered\footnote{In the case of Model C under homogeneous alignment, even though $\arctan\left(\gamma_{Np}\right)$ is near $-90^{\circ}$, only a small rotation is needed to place this value near $+90^{\circ}$, and so are actually much closer than it appears on the plot, which is subject to the limitations of presenting angles on a linear scale.}, displaying little variance between different simulations. 

The column density-polarization fraction joint correlations for a line-of-sight inclined $45^{\circ}$ between the $x$ and $z$ lines-of-sight are presented in Figure \ref{fig:Np45}. The correlation coefficients for the case of 45$^{\circ}$ inclination demonstrate an interesting phenomenon: if one were to consider only the $z$ line-of-sight and then the $x$ line-of-sight, one might be tempted to conclude that as one inclines the line-of-sight between these two extremes, the correlation measures should take intermediate values between the extremes. Indeed, similar intermediate gradation was demonstrated quantitatively in Paper I for the PDFs of polarization fraction and dispersion in polarization angles. However, the 45$^{\circ}$ inclination contradicts this, showing the most positive Pearson coefficients of all three cases, and retaining a positive PCA-implied power-law index under stronger $\beta$ depolarization models (see Figure \ref{fig:Npstats}, middle panel). A comparison between the $z$ line-of-sight homogeneous alignment models and those within the 45$^{\circ}$ inclination case is instructive: one sees echoes of the previously mentioned triangular structure in the 45$^{\circ}$ case, but this structure is eroded at the low column density edge. This may be interpreted by considering the geometry of the colliding flow simulations under the effect of inclining the line-of-sight: the highest column density regions have large contributions from the highest volume density voxels, which are generally compact and have an outsize role in determining the overall polarization in the line-of-sight. Under rotation, these regions should be changed relatively little, especially if their field geometry is affected by self-gravity. On the other hand, the lowest column density regions tend to have fewer compact, high volume density regions, and contributions to polarization tend to be relatively uniform over the line-of-sight, resulting in an overall reduction in polarization fraction due purely to inclination alone. As a result, the intermediate inclination joint correlation between column density and polarization fraction becomes deformed in a non-intuitive way, and in turn results in a joint correlation with surprising characteristics, namely a positive PCA-implied power-law index. Whether this result holds for other types of cloud simulations, particularly those not involving colliding flows, remain to be determined. 

Overall, we have demonstrated that when we use these power-law depolarization models to approximate true heterogeneous alignment, we can obtain a joint correlation between column density and polarization fraction consistent with those obtained by real observations of molecular clouds, such as the BLASTPol observations of Vela C, provided that a suitable choice of parameters is made. Broadly, much of the rotation in the principal components, for both the $x$ and $z$ line-of-sight extremes, happens after the application of power-law depolarization models  with small (in magnitude) $\beta = -1/3$, and the more negative $\beta$ models remain roughly consistent with BLASTPol. A more thorough search through parameter space is necessary to identify a specific $\beta$ or $n_{crit}$ which best matches the BLASTPol data, though one can obtain a rough picture. Insofar as less extreme values (closer to zero) for $\beta$ are more plausible (as the naive interpretation of $\beta$ is related to the average $A_V$ of the cloud) and that the properties of the other observables suggest either a highly inclined line-of-sight or high level of turbulence (Paper I; \citet{CKLFM}), that a $\beta \simeq -1/3$ or smaller in magnitude appears to be consistent with the data. If Figure \ref{fig:ratbeta} is taken at face value, then a relatively weak $\beta = -1/3$ (or even smaller in magnitude) is consistent with an average extinction $A_V \lesssim 10$, perhaps even lower, depending on the model parameters chosen to approximate $\overline{\beta}_{rat}$ as discussed in Sections \ref{subsection:RATDM} and \ref{subsection:approx}. This is roughly consistent with observational information available for Vela C \citep{HVC}, though this glosses over several important caveats. These include the limitations of the treatment outlined in Section \ref{section:PLDM} and that observationally determined $A_V$ are not necessarily the minimum $A_V$ along any sightline, which determines the radiation field that is the alignment mechanism in the radiative torque alignment model.

Importantly, the degeneracy between mean magnetic field orientation and the level of turbulence/magnetization (see Paper I) unfortunately cannot be broken by considering the column density-polarization fraction correlation, but it is encouraging to see that heterogeneous alignment effects may be primarily responsible for the observed anticorrelation, which may allow partial separation of the effects in observationally accessible quantities. On the other hand, the 45$^{\circ}$ inclination results suggest that mixing between different mean magnetic field orientations may have a more complicated role to play than previously expected, though there are choices of power-law depolarization model nevertheless capable of agreement with the BLASTPol observations of Vela C. A more thorough study will necessarily also need to consider some of the uncertainties in the radiative torque alignment modelling discussed in Section \ref{section:PLDM}, as well more realistic forms for depolarization models. We have also not considered the effects of beam convolution, which we chose to neglect in this study as we have found previously in Paper I that it did not alter much of the bulk properties of the joint correlations and would have introduced yet another degree of freedom to explore in our modelling. 

In contrast with the column density-polarization fraction joint correlations, the column density-dispersion in polarization angles joint correlations display little sensitivity to our power-law depolarization models. This is strikingly evident in Figure \ref{fig:NSPL}. For quantitative comparison, we present $\arctan(\gamma_{NS}$ and $\rho_{NS}$ in Figure \ref{fig:NSstats}. As in Paper I, this correlation when viewed along the $z$ line-of-sight displays a moderate positive correlation, which is very steep; much of this relationship can be explained by the lower variance in column density in comparison with the $x$ line-of-sight, which displays uncorrelated behaviour similar to that seen in the BLASTPol observations of Vela C. As established in Section \ref{section:pl1D}, the populations of the dispersion in polarization angles are affected by the power-law depolarization models, but crucially, the effect on the joint correlation appears to be minimal. Some depletion of the highest $\mathcal{S}$ regions (the tops of the contours) is evident at the strongest depolarization models, but even these have a weak overall effect on the principal components, being rotated or dilated very little. This provides further evidence that the dispersion in polarization angles and column density structures are generically uncorrelated with each other to any significant degree. 

\subsubsection{Column Density-Polarization Fraction Correlations and Line-of-Sight Structure under Heterogeneous Alignment} \label{subsubsection:coldenlos}

These properties of the column density-polarization fraction joint correlation under power-law depolarization models are consistent with arguments previously advanced in Paper I regarding the differences between the $z$ and $x$ line-of-sight, and therefore the role of magnetic organization with respect to the observer. It was observed that the $x$ line-of-sight shared some characteristics of high (apparent) magnetic field disorder with the more turbulent/less magnetized simulations when viewed from the $z$ line-of-sight: namely, that the overall level of the dispersion in polarization angles $\mu_G(S)$, the geometric standard deviation of the polarization fraction $\sigma_G(p)$, and the PCA-implied power-law index of the joint correlation between them, of the higher turbulence simulations are driven closer to the $x$ line-of-sight values as the relative importance of turbulence increases (see Figures \ref{fig:1DPL}, \ref{fig:means}, and \ref{fig:Spstats}; also, Figures 4, 5, and 6 in Paper I). The response of the column density-polarization fraction joint correlation to power-law depolarization models appears to also have a graded response to the level of turbulence/magnetization (Figure \ref{fig:Npstats}), and again, in the context of this response, the $x$ line-of-sight seems to behave as if it is an extremely magnetically disordered configuration. 

The response of each simulation to the power-law depolarization models is closely related to line-of-sight density structure and magnetic organization. Within the $z$ line-of-sight, less turbulent conditions result in more narrow volume density distributions within a typical sightline, with extreme variations from the mean volume density less common and therefore relatively less important.\footnote{Note that this refers not only to volume density enhancements but also to volume density reductions as well; lower levels of turbulence result in more uniform density distributions.} As the polarization response of higher volume density regions is more attenuated than lower ones by a power-law depolarization model, the high contributions of these regions to column density will not be matched by a high contribution to polarized intensity (and therefore polarization fraction); provided that high volume density regions are more common within the high column density regions, then the mechanism by which a column density-polarization fraction anticorrelation is induced by power-law depolarization models becomes clear.\footnote{This is not to say that the less turbulent conditions are not affected by the power-law depolarization models at all - merely that the relative response of the lower volume density regions, even if they are larger than $n_{crit}$, will not be as large as the response of the highest volume density regions, which are much more common in higher turbulence conditions relative to low turbulence conditions.} This also explains the graded response of the simulations: since lower turbulence regions have fewer extreme density variations and therefore have fewer of the most strongly affected high volume density regions, then lower turbulence simulations should be affected less than the higher turbulence simulations by our power-law depolarization models, which is indeed the case in this line-of-sight.

This argument is, however, not completely satisfactory when applied to the $x$ line-of-sight, which fails to display this gradation with level of turbulence. Within the $x$ line-of-sight, the base state is that of already very low polarization fraction - indeed, were the magnetic field perfectly aligned with the observer's line-of-sight with no variations into the plane-of-sky, the polarized intensity would be identically zero everywhere. As a result, variations in this line-of-sight must be considered relative to a state of zero, not maximal, polarization. Unlike maximal polarization conditions, a state of zero polarization fraction is easy to disrupt - a single region bent out of the line-of-sight orientation is sufficient to induce a polarization response. As a result, even very low turbulence levels are sufficient to break these conditions, and cancellation within the line-of-sight becomes far more important; but, at the same time, even very high turbulence conditions are insufficient to break out of the dominant low polarization level, \textit{unless} it is strong enough to change the orientation of the mean magnetic field (in which case the relative orientation of the observer and the magnetic field has changed). As a result, it is not just consistent but a requirement that turbulence level should have very little effect on the response to heterogeneous alignment in the $x$ line-of-sight, and even a weak amount of depolarization due to grain alignment physics should be sufficient to induce a negative correlation for precisely the same reasons outlined above in the $z$ line-of-sight case. 

The response of the Pearson correlation coefficient, $\rho_{Np}$, is somewhat simpler to interpret. In a more disordered context, the larger variations seen within the populations may tend to drive $\rho_{Np}$ closer to zero, as the presumed underlying correlation is disrupted by these variations. This hypothesis is supported by the gradation within the $z$ line-of-sight of $\rho_{Np}$ with turbulence level (see Figure \ref{fig:Npstats}), with more ordered simulations (Models C and D) more anticorrelated than the more turbulent/less magnetically ordered ones (Models A and B). That being said, the $x$ line-of-sight $\rho_{Np}$ in comparison does not display this effect in an especially pronounced way. Instead, the high state of magnetic disorder in the $x$ line-of-sight could explain the relative lack of general trend with respect to level of turbulence within the simulations in the $x$ line-of-sight, while displaying the overall trend of steeper $\beta$ anticorrelating $N$ and $p$ mores strongly. 

\subsection{The Polarimetric Joint Correlation} \label{section:plSp}

\begin{figure*}
\centering
\vspace{-2mm}
\includegraphics[height=0.46\textheight, width=\textwidth]{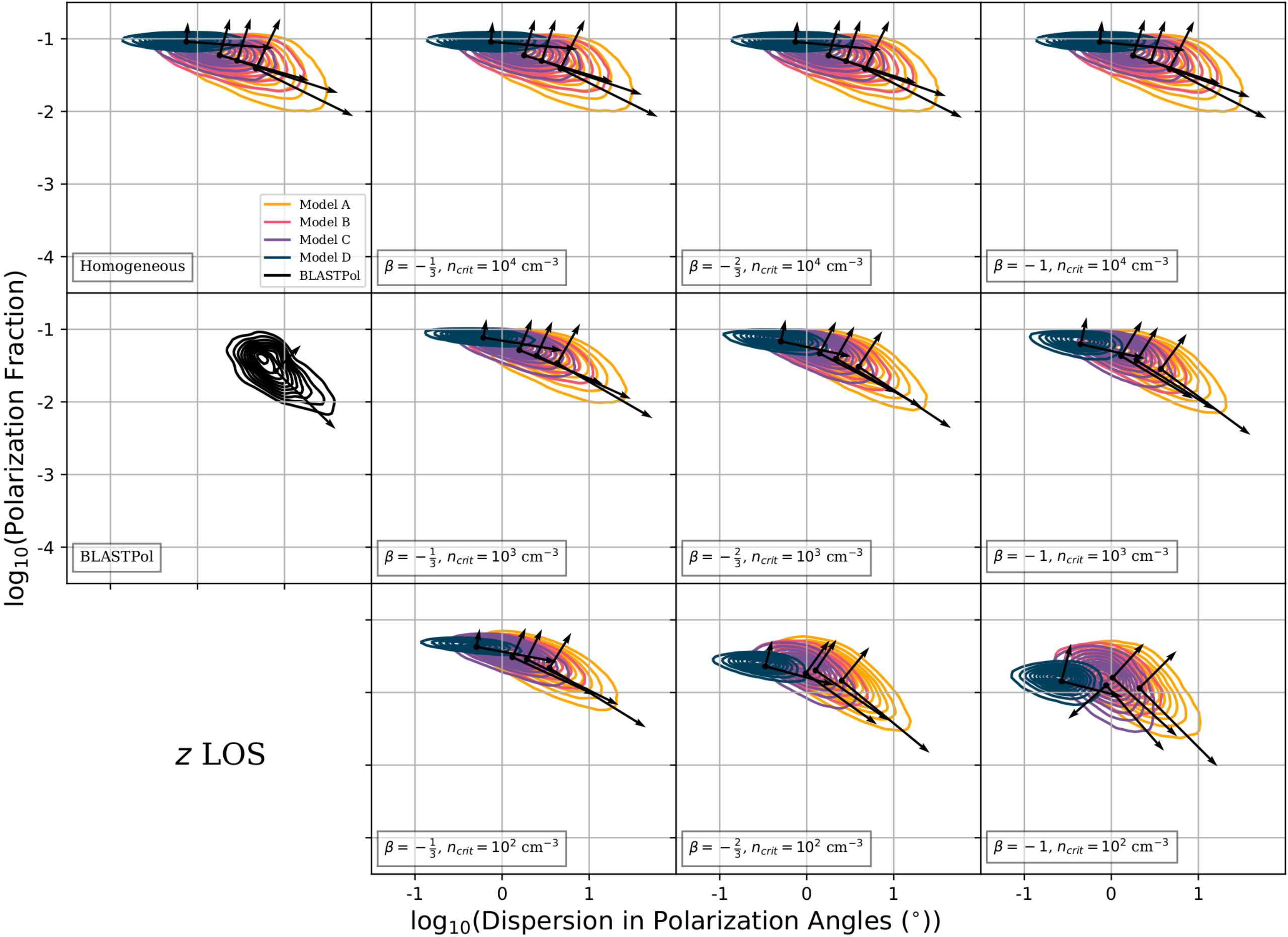}\\
\vspace{1mm}
\includegraphics[height=0.46\textheight, width=\textwidth]{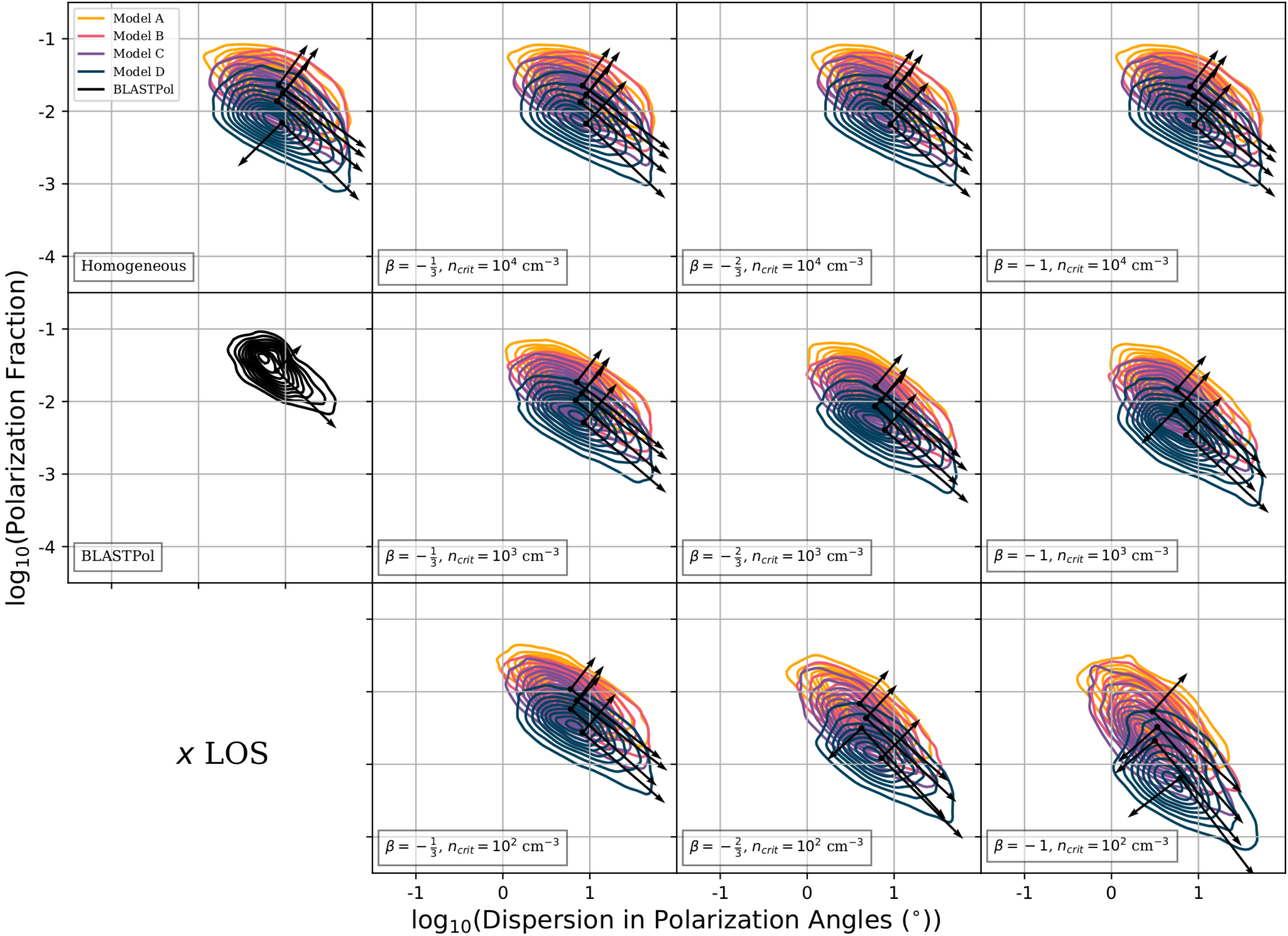}
\vspace{-5mm}
\caption{Same as Figure \ref{fig:NpPL}, but for the correlation between dispersion in polarization angles and polarization fraction.}
\label{fig:SpPL}
\end{figure*}

\begin{figure}
\centering
\includegraphics[width=\columnwidth]{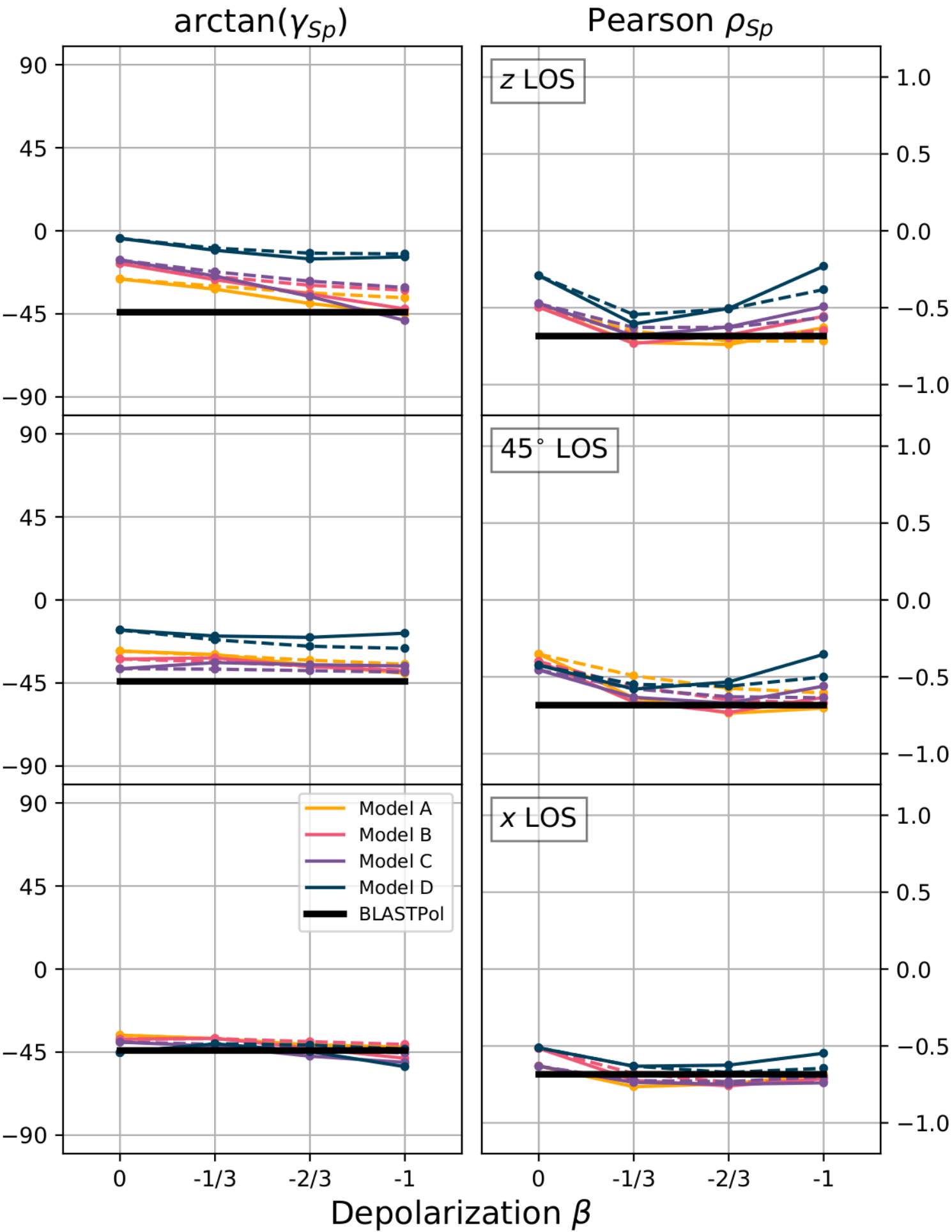}
\vspace{-4mm}
\caption{Same as Figure \ref{fig:Npstats}, but for the correlation between the dispersion in polarization angles and polarization fraction.}
\label{fig:Spstats}
\end{figure}

In Paper I, the correlation between $p$ and $\mathcal{S}$ (termed the polarimetric joint correlation) was found to be generically negatively correlated for both simulations considered in the study, a result consistent with previously reported observations \citep{PXIX}. In the $x$ line-of-sight, the correlation was both robust and near the BLASTPol value, and given that the 1D PDFs of $p$ and $\mathcal{S}$ in the $x$ line-of-sight were consistent with BLASTPol as well, this was taken as evidence suggesting that the mean magnetic field in Vela C is oriented mostly parallel to our line-of-sight. This assertion has been strengthened by recent work that provides an alternative means of estimating the mean magnetic field orientation using dust polarization information \citep{CKLFM}. At the same time, consideration of the two simulations when viewed from the $z$ line-of-sight revealed a dependence of this joint correlation on the level of turbulence present. Taken together, the evidence suggested that strong anticorrelation of $p$ and $\mathcal{S}$ with a power-law index of -1 was a signature of high apparent magnetic field disorder, which either a line-of-sight nearly parallel to the mean magnetic field or high level of turbulence/low level of magnetization drive towards. It is important to examine whether adopting our power-law depolarization model renders these previous assertions invalid. 

Because we have adopted a larger selection of simulations in this study, we first consider the synthetic observations under homogeneous alignment, presented in the top left subpanels of both panels of Figure \ref{fig:SpPL}. Quantitatively, $\arctan(\gamma_{Sp})$ and $\rho_{Sp}$ are presented in Figure \ref{fig:Spstats}. In both the $x$ and $z$ lines-of-sight, the new simulations provide further evidence for the conclusions drawn in Paper I. Within the $x$ line-of-sight, there appears to be little to no difference in the polarimetric joint correlations between the different simulations; within the $z$ line-of-sight, there is a gradation between the different simulations, with more turbulent simulations presenting a steeper PCA-implied power-law index. One piece of evidence supporting the conclusion that Vela C may have a highly inclined mean magnetic field was that, even though higher turbulence drives the polarimetric joint correlation to steeper anticorrelation, this effect was not sufficient to bring it into agreement with BLASTPol. We find that adding an even higher turbulence simulation (Model A) still fails to achieve agreement, strengthening this conclusion. 

The power-law depolarization models are presented in the other subpanels in Figure \ref{fig:SpPL}, in the same order as presented in Figures \ref{fig:NpPL} and \ref{fig:NSPL}. Remarkably - despite the significant changes to the distributions of the polarization fraction and dispersion in polarization angles alone under power-law depolarization models, discussed in Section \ref{section:pl1D} - the joint correlations in the $x$ line-of-sight (bottom panel of Figure \ref{fig:SpPL}) display exceptionally little variation from the homogeneous alignment case. This is not at all expected \textit{a priori}, especially considering that changes to the grain alignment model resulted in significant changes to the column density-polarization fraction joint correlations, and that the effects on the individual observables were not equal in magnitude for each simulation (see Figure \ref{fig:means}). However, given that the dispersion in polarization angles and column density remain generally uncorrelated (see Section \ref{section:plCD}), it appears to be the case that despite an unequal overall reduction in $\mu_G(p)$ and $\mu_G(\mathcal{S})$ and the inducement of a column density-polarization fraction correlation, the relationship between $\mathcal{S}$ and $p$ appears to be preserved under heterogeneous alignment.

In the $z$ line-of-sight (top panel of Figure \ref{fig:SpPL}) the gradation in the steepness of the polarimetric joint correlations also appears to be preserved under power-law depolarization models, though with less consistency. The higher turbulence simulations tend to be more susceptible to the tendency of a stronger power-law depolarization model to draw the PCA-implied power-law index towards $-1$ ($-45^{\circ}$ relative to horizontal.) This suggests that there may be some additional degeneracy that heterogeneous alignment adds when considering the role of the level of turbulence/magnetization and the orientation of the mean magnetic field: while turbulence/magnetization alone is not quite sufficient to mimic the $x$ line-of-sight, maximally disordered conditions, and (considering the effects of power-law depolarization models on the least turbulent simulation, Model D) while our heterogeneous alignment models alone are also insufficient, taken together it is possible to reproduce a steep PCA-implied power-law index like that found in the BLASTPol observations of Vela C. This indicates that the PCA-implied power-law index for the polarimetric joint correlation is a useful signature of magnetic structure that is unfortunately not completely separable from grain alignment physics.

The insensitivity of the PCA-implied power-law indices of the polarimetric joint correlations in the $x$ line-of-sight and the preservation of the gradation with turbulence/magnetization in the $z$ lines-of-sight points again to the unified picture of magnetic field disorder having the same effects, regardless of its origin (i.e. whether it arises due to turbulence or orientation of the mean magnetic field). This connection has been explained in terms of the role that cancellation within the line-of-sight plays: given overall inclination characteristics of the mean magnetic field, and that the distributions of volume density (and, by proxy, grain alignment efficiency) remain relatively fixed, those regions with high variability of the mean magnetic field within the line-of-sight tend to have the lowest polarization levels relative to the population, and in turn these regions tend to be those with outsize variability in polarization angle between adjacent lines-of-sight (i.e. high $\mathcal{S}$). Only by suppressing magnetic disorder, such as Model D viewed along the $z$ line-of-sight, can this anticorrelation be suppressed. 

Lastly, we note that the elimination of the filamentary high-$\mathcal{S}$ structures by strong depolarization power-law models, as discussed in Section \ref{section:pl1D}, appears to have a minimal effect on the joint correlations, suggests that these regions likely do not have a large role to play in the overall joint correlations. Overall, it is likely that they have a different origin than the rest of the quiescent $\mathcal{S}$ population (i.e. the regions of $\mathcal{S}$ which are not part of the filamentary high-$\mathcal{S}$ regions), consistent with their peculiar filamentary structure. 

\section{Conclusions} \label{section:conc}

To summarize:  

\begin{enumerate}

\item{Proceeding from the basic equations for the linear Stokes parameters arising from optically thin polarized thermal emission from magnetically aligned dust grains (Section \ref{section:hetalign}), we introduce a generic method for incorporating the effects of heterogeneous grain alignment based on the radiative torque theory \citep{CL1}, in which the alignment efficiency may depend on local conditions (Section \ref{section:PLDM}). We present a form for the grain alignment efficiency in the form of a basic power-law in local gas density (Section \ref{subsection:PLDM}) and justify it using results derived from radiative torque alignment theory (Sections \ref{subsection:RATDM}, \ref{subsection:approx}, and \ref{subsection:justify}).}

\item{Using this heterogeneous alignment prescription, we make synthetic observations of an expanded set of four simulations beyond those considered in Paper I and present them in Section \ref{section:plresults}. We first consider the probability density functions (PDFs) of the polarization fraction $p$ and the dispersion in polarization angles $\mathcal{S}$ (Figure \ref{fig:1DPL}). Considering the naive expectation that the $p$ populations should be affected more than the $\mathcal{S}$ populations by heterogeneous grain alignment, we find that this expectation does not hold up when considering the PDFs at face value. As measured by the geometric means (Figure \ref{fig:means}), the $p$ population within the $z$ line-of-sight (in which the mean magnetic field is perpendicular to the line-of-sight) shows a graded response to heterogeneous alignment that depends on turbulence/magnetization, in which higher turbulence/lower magnetization tends to drive the $p$ population down more. Within the $x$ line-of-sight (in which the mean magnetic field is parallel to the line-of-sight) the $p$ population does not have the same clear response to turbulence/magnetization. The $\mathcal{S}$ populations seem to display the opposite response within each line-of-sight - in the $z$ line-of-sight, the $\mathcal{S}$ populations respond in nearly the same way, but the $\mathcal{S}$ populations within the $x$ line-of-sight show a graded response to turbulence/magnetization. Altogether, the $\mathcal{S}$ populations are affected more than naively expected (often on the same order as the effect on the $p$ populations) and display interesting behaviour with respect to the apparent magnetic structure (line-of-sight and turbulence/magnetization.) }

\item{We then examine the $\mathcal{S}$ populations more closely, and note that the bulk of the effect of heterogeneous alignment on the $\mathcal{S}$ PDFs is at the high end of the population near the characteristic value, $\pi/\sqrt{12}$ \citep{PXIX}. Visually (Figure \ref{fig:Scompare}) the filamentary high-$\mathcal{S}$ structures tend to be strongly attenuated with a strong power-law choice for heterogeneous alignment. This suggests two things. First, that the majority of the effects on the $\mathcal{S}$ populations by heterogeneous alignment might be confined to these unique regions, and that the bulk of the distribution is not changed much by heterogeneous alignment. This recovers the naive expectation mentioned previously, which would then hold for most polarization observations, save for these unique filamentary high-$\mathcal{S}$ regions, and makes $\mu_G(\mathcal{S})$ a very important quantity to measure. Second, the attenuation behaviour with more negative $\beta$ (stronger dependence of polarization efficiency on volume density; see Equation \eqref{p0powerlaw}) suggests that these unique high-$\mathcal{S}$ features may originate from higher density regions within the molecular cloud, but those that are not self-gravitating (and hence the lack of $N$ vs. $\mathcal{S}$ correlation). These regions are generally present within all lines-of-sight in any oblique magnetized shock \citep{CLKF}, which may be general phenomena within star-forming regions.}

\item{When we examine the correlations involving column density, we find that the PCA-implied power-law index - corresponding to the slope of the correlation - is readily rotated to values consistent with the observational evidence when power-law depolarization models are introduced. The Pearson correlation coefficient also tends to be made more negative, which again is more consistent with the observational evidence (which shows a moderately strong anticorrelation.) The correlation in the $z$ line-of-sight displays a graded response with respect to the turbulence/magnetization, whereas in the $x$ line-of-sight there is little response to the same. This is understood by considering the role that turbulent perturbations play within these two regimes of magnetic structure: on the one hand, the $z$ line-of-sight (nearly perpendicular to the mean magnetic field) is apparently magnetically well-ordered, and in the presence of stronger turbulence, the likelihood of high density voxels within the line-of-sight is increased, making the more turbulent simulations more susceptible to power-law depolarization models. On the other hand, within the $x$ line-of-sight (nearly parallel to the mean magnetic field), an unperturbed line-of-sight would have nearly zero polarization, and as a result, any level of turbulence can introduce a small, non-zero level of polarization. Finally, unlike the correlation between column density and polarization fraction, the correlation between column density and dispersion in polarization angles appears generally uncorrelated across our heterogeneous alignment models, simulations, and lines-of-sight.}

\item{We also present the curious case of the line-of-sight inclined 45 degrees between the $x$ and $z$ lines-of-sight, in which it is demonstrated that it is possible to induce a positive correlation between column density and polarization fraction under homogeneous alignment (see Figure \ref{fig:Np45}). We argue that this is a result of the distinct effects inclination of the mean magnetic field might have within a turbulent region subject to self-gravity: that the densest regions, being self-gravitating and compact, should have Stokes parameters which do not vary much when rotated, but the bulk, lower to medium volume density regions will be affected by inclination reductions to the polarization fraction. As a proof-of-concept this case demonstrates that the overall effects of magnetic structure (in this case, mean magnetic field orientation) may have different apparent effects on different observed statistical features in the MC. Regardless, the application of heterogeneous alignment models converts this positive polarization fraction-column density correlation to a negative one under suitable parameter choices.}

\item{Finally we return to the polarimetric joint correlation (between $p$ and $\mathcal{S}$). We find that despite the overall effects on the populations of $p$ and $\mathcal{S}$ (Section \ref{section:pl1D}) and the inducement of an anticorrelation between column density and $p$, the polarimetric joint correlation is relatively weakly affected by heterogeneous alignment, being seen only within the $z$ line-of-sight and degenerate with the effects of turbulence/magnetization. There is nearly no effect on the correlation within the $x$ line-of-sight, which points to the significance of apparent disorder regardless of origin (i.e. turbulence or inclination.) Ultimately we find that the introduction of heterogeneous alignment does not destroy the good agreement found in previously in Paper I between the $p$ vs. $\mathcal{S}$ correlations of BLASTPol and the synthetic observations.}

\item{Taken altogether, the evidence from each of Sections \ref{section:pl1D}, \ref{section:plCD}, and \ref{section:plSp} suggests that, under reasonable choices for the parameters in our power-law heterogeneous grain alignment prescription, it is possible to induce a negative correlation between column density and polarization fraction without destroying the good agreement on the $p$ vs. $\mathcal{S}$ correlation between the BLASTPol results and the synthetic observations when viewed from lines-of-sight very close to the mean magnetic field, as reported in Paper I. If the analysis proceeding from radiative torque theory presented in Section \ref{section:PLDM} is taken at face value, this suggests that Vela C is a molecular cloud with a magnetic field pointing nearly parallel to the line-of-sight of the observer, and that the less extreme, small magnitude $\beta \lesssim -1/3$ necessary to induce a negative correlation corresponds to an average $A_V \lesssim 10$ or perhaps lower depending on specific model conditions} (see Equations \eqref{dlogpdlogn} and \eqref{scale}), which is roughly consistent with observational evidence \citep{HVC}. However, this must be understood with several caveats, including the limitations of the modelling in Section \ref{section:PLDM}.

\end{enumerate}

\section*{Acknowledgements}

PKK is supported by a Livermore Graduate Scholarship at Lawrence Livermore National Laboratory, and acknowledges the Jefferson Scholars Foundation for additional support through a graduate fellowship and NRAO through a SOS (Student Observing Support) award. CYC, LF, and ZYL acknowledge support from NSF AST-1815784 and NASA 80NNSC18K0481. ZYL is supported in part by NASA 80NSSC18K1095 and NNX14AB38G and NSF AST-1716259. This work has been partially performed in support of developing analysis tools for the BLAST collaboration. BLAST is supported by NASA under award numbers NNX13AE50G and 80NSSC18K0481. Part of this work was performed under the auspices of the Department of Energy by Lawrence Livermore National Laboratory under Contract DE-AC52-07NA27344. LLNL-JRNL-764043.

\bibliographystyle{mnras}
\bibliography{references} 

\appendix

\bsp	
\label{lastpage}
\end{document}